\documentclass{aa}

\usepackage{linenoaa}
\usepackage{hyperref}
\usepackage{siunitx}
\usepackage{booktabs}
\usepackage{tabularx}
\usepackage{array}
\usepackage{graphicx}
\usepackage{subfig}
\usepackage{afterpage}
\usepackage{amsmath}
\usepackage{upgreek}

\usepackage{txfonts}

\usepackage{hyperref}

\begin{document} 

   \title{Post-perihelion dust activity and a candidate icy-grain halo in long-period comet C/2023 A3 (Tsuchinshan–ATLAS)}

   \author{Bin Liu
          \inst{1}
          \and
          Xiaoping Zhang
          \inst{1}
                    \and
          Xiaodong Liu\inst{2}
                    \and
          Man-To Hui\inst{3}
          }
    \institute{
    State Key Laboratory of Lunar and Planetary Sciences, Macau University of Science and Technology, Taipa, Macau 999078, People's Republic of China
    \and
    School of Aeronautics and Astronautics, Shenzhen Campus of Sun Yat-sen University, Shenzhen, Guangdong 518107, People's Republic of China
    \and
    Shanghai Astronomical Observatory, Chinese Academy of Sciences, No. 80 Rd Nandan, Shanghai 200030, Mainland China\\
    \email{xpzhangnju@gmail.com}
    }

   \date{}
\titlerunning{Post-perihelion icy grains in C/2023 A3}

\authorrunning{Liu et al.}

\abstract
{We analyze the post-perihelion coma of the Great Comet of 2024, C/2023 A3 (Tsuchinshan–ATLAS), to characterize its dust activity and assess the contribution of sublimating icy grains to a candidate near-nucleus halo. Using archival Zwicky Transient Facility images obtained between 15 February and 26 May 2025, we performed aperture photometry within a fixed projected radius of 40,000 km and derived azimuthally averaged surface-brightness profiles from annular photometry. Over the observed interval, the comet faded from 12.86 to 14.77 mag as its heliocentric distance increased from 2.64 to 3.90 au, although two short-lived brightenings were superimposed on the secular decline. The effective dust cross-section decreased by about a factor of 2.4 and followed $C_{\rm e} \propto r_{\rm h}^{-3.1}$, while the dust mass production rate scaled as $\dot{M}_{\rm d} \propto r_{\rm h}^{-3.41}$. These dependences are compatible with a contribution from water-ice-driven activity. Extrapolating the fitted post-perihelion relation back to perihelion gives a model-dependent cumulative dust mass loss of $3.0\times10^{11}$ kg, corresponding to a globally averaged erosion depth of $\sim 1.4$ m for an assumed nucleus radius of 5.9 km. The coma profiles are shallower than the canonical steady-state expectation, indicating radial evolution in the dust distribution or scattering properties, with delayed aggregate fragmentation as one possible contributor. In addition, a persistent central excess, spatially coincident with a less-red inner-coma color, supports the presence of an additional near-nucleus component consistent with an icy-grain halo. Among the cases explored, icy grains with radii of $\sim 10~\mu$m and a carbon fraction of $\sim 0.1 \%$ yield model halo radii comparable to the observed halo scale during the enhanced-activity interval between 21 and 25 February 2025, with model grain lifetimes ranging from $2.8\times10^5$ to $3.2\times10^5$ s. Although no single model with both grain size and composition held fixed reproduces the full halo evolution, an increase in the carbon fraction of the released icy-grain population provides one possible explanation for the broad halo-radius trend and may reflect progressive refractory enrichment of the evolving near-surface material.}

\keywords{comets: individual: 2023 A3 (Tsuchinshan–ATLAS) / methods: observational / methods: numerical / techniques: photometric}

\maketitle
\nolinenumbers
\section{Introduction}
Comets are widely regarded as primitive bodies that retain a record of material inherited from the early Solar System \citep{betzler2012nonextensive}. As a comet approaches the inner Solar System, solar heating progressively drives the sublimation of volatile ices from its near-surface layers. The escaping gas exerts drag on surface dust, lifting it from the nucleus into the surrounding space and thereby forming a coma composed of gas and solid particles \citep{lejoly2022radial}. Because the dust both preserves relatively primitive mineral and organic constituents of the nucleus and records the release, transport, and subsequent evolution of cometary material, the coma provides a link between nucleus properties and remote observations, as well as a diagnostic of nucleus activity, dust properties, and their origins and evolutionary pathways \citep{biver2022chemistry}.

\begin{figure}
\centering
    \includegraphics[width=\columnwidth]{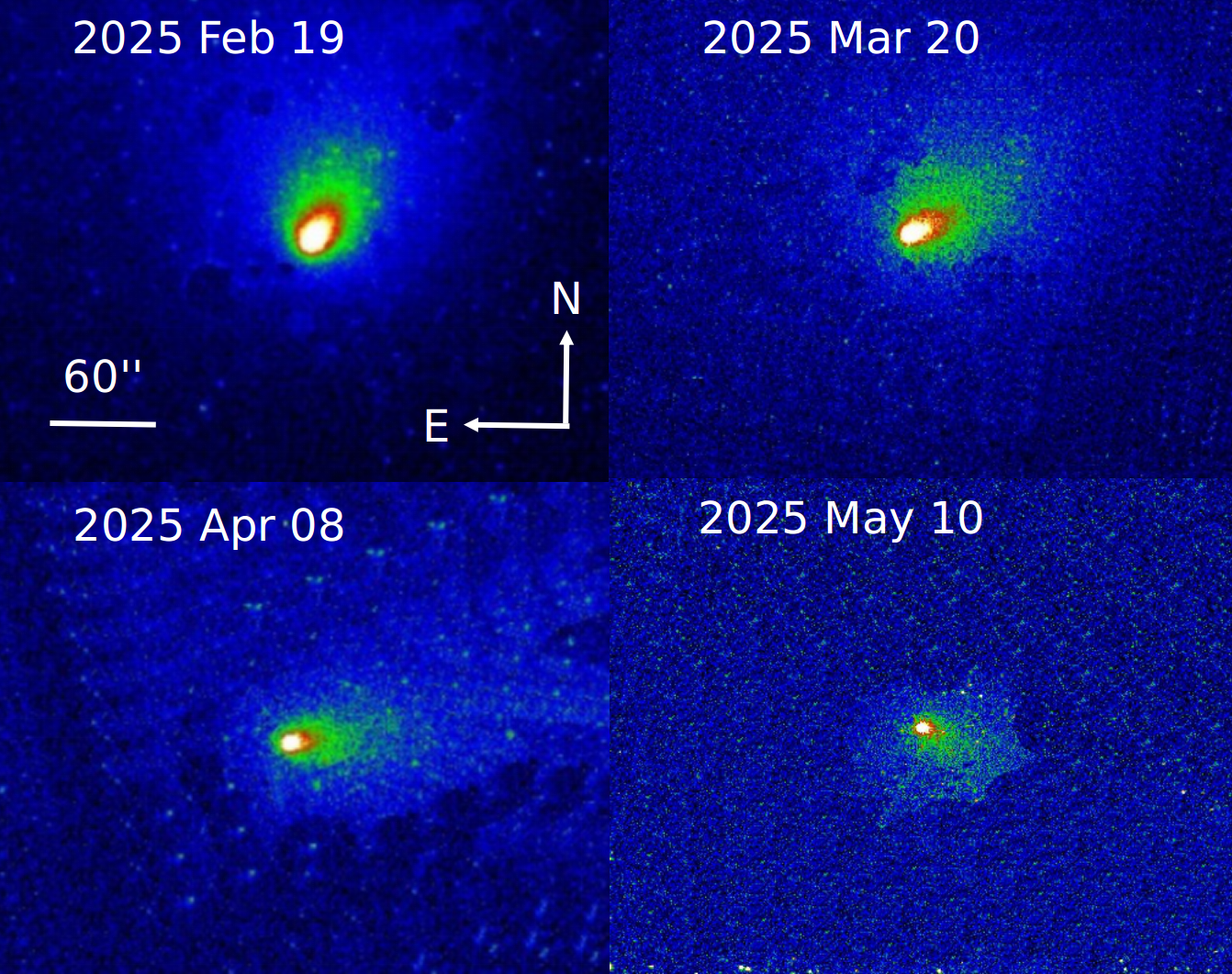}
    \caption{Representative composite images of C/2023 A3 at four post-perihelion epochs. The observing date is shown in the upper left of each panel. The scale bar and orientation are indicated in the upper-left panel.}
    \label{ObsImag}
\end{figure}

Rosetta measurements have been interpreted within a pebble-based picture of cometary nuclei. Compact particles detected at comet 67P/Churyumov--Gerasimenko were interpreted as millimetre- to centimetre-sized aggregates of ice and refractory material, representing possible cometary building blocks \citep{fulle2016comet}. Subsequent GIADA measurements constrained the ice content and microporosity of these icy pebbles \citep{fulle2017dust}. Extending this picture to the nucleus scale, \citet{blum2017evidence} proposed that 67P formed through the gentle gravitational collapse of a bound clump of millimetre-sized dust aggregates mixed with microscopic ice particles. At the scale of an individual pebble, \citet{fulle2019comets} considered sublimation and gas diffusion within a porous ice--dust aggregate. In their model, gas diffusion from the pebble interior generates a steep density gradient near the surface, which may fragment an illuminated pebble and release smaller, sub-pebble particles. A related proposed activity scenario invokes Water-Enriched Bodies (WEBs), approximately metre-scale concentrations of water-rich pebbles embedded in more water-poor and CO$_2$-rich material. In this framework, WEBs may act as localized sources of water-driven dust release \citep{attree2024localized}.	

Once these particles enter the coma, their initial evolution is controlled primarily by gas drag. In the near-nucleus region, grains may continue to gain momentum from gas drag, but they subsequently evolve more independently as the gas density decreases and gas-grain coupling weakens \citep{lejoly2022radial}. The inner coma therefore constitutes a transitional environment in which both dust dynamics and intrinsic grain properties play major roles. For grains that contain volatile ice, sublimation is particularly important. Ice loss modifies both the grain mass and its effective scattering cross-section, while anisotropic outgassing from the grain surface or interior can generate recoil forces that further perturb the grain trajectory \citep{agarwal2024dust}. Meanwhile, sublimation-driven pressure build-up may trigger fragmentation, causing larger aggregates to break into smaller fragments and thereby reshaping both the grain-size distribution and the surface brightness morphology of the inner coma \citep{jewitt1987surface,markkanen2020thermophysical}.

Imaging photometry provides an effective means of tracing grain evolution within the coma. For an optically thin coma sustained by a steady dust source, the surface brightness on the sky plane is expected to follow $I(\rho)\propto \rho^{-1}$, provided that grains expand outward at nearly constant velocity and maintain broadly unchanged scattering properties \citep{jewitt1987surface}. Departures from this canonical behavior generally indicate that the grains’ albedo, size distribution, or dynamical state varies with projected nucleocentric distance \citep{jewitt1987surface}. In the case of fragmentation, the breakup of large aggregates during outflow can increase the total scattering cross-section at larger distances and thus flatten the radial brightness profile, although the detailed observational signature depends on grain composition and on the relative importance of concurrent processes \citep{stevenson2012near}. Similarly, if the coma contains ice-bearing grains, the near-nucleus region may exhibit additional brightness excess, again producing deviations from the simple $I(\rho)\propto \rho^{-1}$ profile \citep{stevenson2012near}. Comparable departures from the canonical radial profile have been reported in several comets. Fragmentation was invoked to explain the profile evolution of the Jupiter-family comet (JFC) 45P/Honda-Mrkos-Pajdušáková \citep{lejoly2022radial}, while near-nucleus signatures associated with icy-grain halos were identified in the long-period comet C/1996 B2 (Hyakutake) \citep{harris1997evidence} as well as in the JFCs 17P/Holmes \citep{stevenson2012near} and 103P/Hartley 2 \citep{meech2011epoxi}.

Long-period comet C/2023 A3 (Tsuchinshan--ATLAS), hereafter comet A3, provides a case for examining these issues. Comet A3 exhibited pronounced dust activity around perihelion in 2024, accompanied by prominent coma and tail structures, including shells, dark lanes aligned with the tail axis, and an anti-tail \citep{moreno2025dust,moreno2025dustA,liu2025great}. Dust tail modeling has further suggested that the dust production rate, size distribution, and minimum grain size all evolved significantly near perihelion, while high-spatial-resolution morphological analyses have indicated that some of these structures may be associated with short-lived, intense emission events dominated by small particles \citep{moreno2025dustA}. In addition, pre-perihelion observations revealed comparatively high $Af\rho$ values, implying a substantial dust-continuum contribution during at least part of the active phase \citep{cambianica2025pre}. The polarimetric studies of comet A3 have shown that the measured polarization exhibits a high maximum degree of linear polarization \citep{lim2025optical}. This behavior is consistent with a dust population dominated by relatively porous aggregates, since more compact aggregates generally produce stronger depolarization \citep{kolokolova2012polarization}.
%%260330

Taken together, these results suggest that the pre- and post-perihelion phases of comet A3 provide a context for investigating its physical properties. Moreover, its small perigee distance of about 0.47 au provided a favorable opportunity for ground-based observations to constrain them without the need for spacecraft measurements. However, comet A3's physical state around perihelion remains open to debate. \citet{sekanina2024inevitable} argued that the brightness anomaly, together with the unexpectedly high apparent water-production rate reported by \citet{ahuja2024molecular}, points to progressive fragmentation of the nucleus prior to perihelion, whereas \citet{liu2025great} contended that comet A3 remained close to a steady state over their observing interval and that the nucleus was stable against tidal disruption, sublimation-driven failure, and rotational instability. In this context, unresolved icy grains or ice-bearing aggregates in the near-nucleus coma could provide an extended source of water and may contribute to the brightness anomaly and the elevated apparent water-production rate.

\begin{figure*}[!t]
	
	\begin{minipage}[t]{0.7\textwidth}
		\vspace{0pt}
		\centering
		\includegraphics[width=0.49\linewidth]{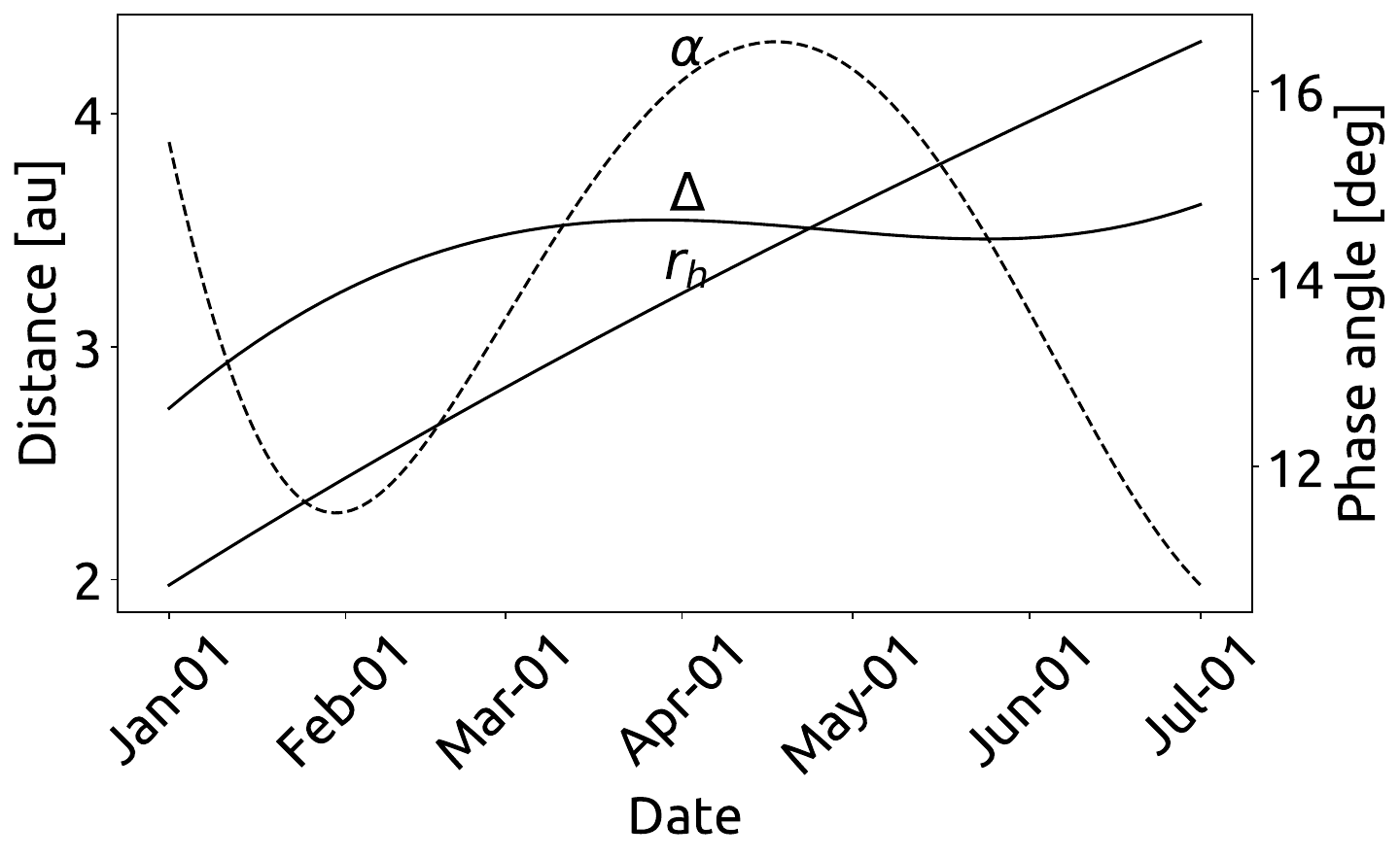}
		\hfill
		\includegraphics[width=0.49\linewidth]{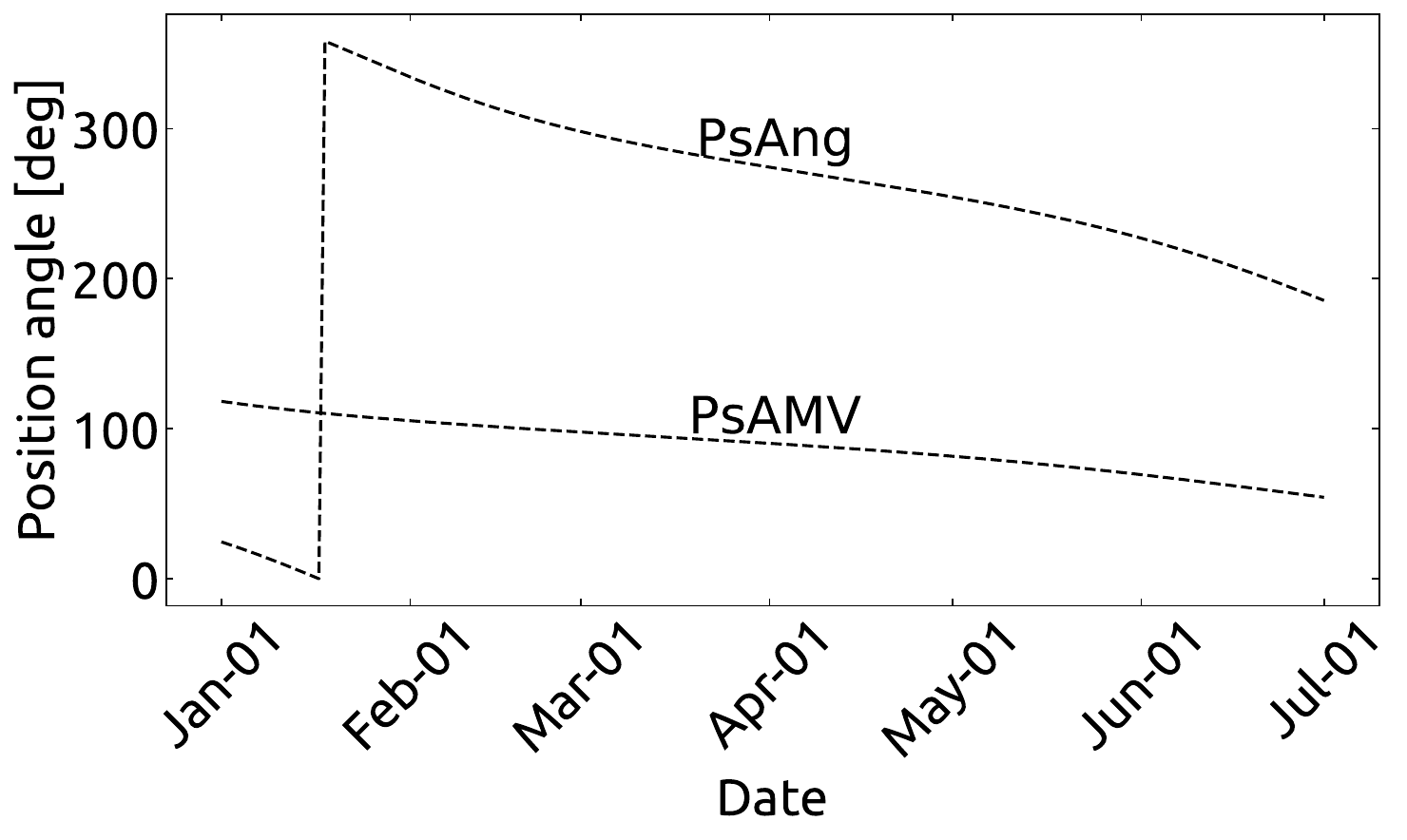}
	\end{minipage}
	\hfill
	\begin{minipage}[t]{0.29\textwidth}
		\vspace{0pt}
		\caption{Observing geometries of comet C/2023 A3. The left panel shows the
			temporal evolution of the phase angle $\alpha$, geocentric distance $\Delta$,
			and heliocentric distance $r_{\rm h}$. In the right panel, PsAng and PsAMV
			indicate the position angles of the projected Sun--A3 direction and the
			negative heliocentric velocity vector, respectively.}
		\label{geometry_A3}
	\end{minipage}
	
\end{figure*}

Motivated by the considerations above, this work investigates the post-perihelion activity of comet A3. In Sect.~\ref{Data}, we describe the observational data and the associated reduction and photometric procedures. In Sect.~\ref{Results}, we present the photometric results, including the evolution of the dust activity and the radial surface-brightness profiles of the coma. In Sect.~\ref{discussion}, we discuss the implications of these results for the post-perihelion dust loss, the origin of the shallow surface-brightness profiles, and the possible presence of an icy-grain halo. Finally, Sect.~\ref{CONCLUSIONS} summarizes our main conclusions.

\section{Data}
\label{Data}
To investigate the morphological evolution of C/2023 A3, we retrieved archival images from the Zwicky Transient Facility (ZTF; \citealt{graham2019zwicky}), a time-domain survey conducted at Palomar Observatory with a 48-inch Schmidt telescope and a 600-megapixel camera, providing a field of view of 47 deg$^{2}$ and a pixel scale of \(1\arcsec\) pixel$^{-1}$. We selected \(r\)-band images obtained on 20 distinct dates between 15 February and 26 May 2025 for analysis, all of which were acquired during the post-perihelion phase, when C/2023 A3 remained active. To enable a uniform comparison of the coma morphology, we extracted a \(6\arcmin \times 6\arcmin\) subframe centered on the nucleus from each image and removed background stars and cosmic-ray hits within the cropped region. These reduction steps were carried out using AstroArt \citep{nicolini2003astroart} and SAOImage DS9 \citep{joye2003new}. Representative processed images, selected to illustrate the morphological evolution of the coma over the observing interval, are shown in Fig.~\ref{ObsImag}.

Figure~\ref{geometry_A3} shows the corresponding observing geometry over the selected interval. Both the heliocentric and geocentric distances remained within \(\sim 2\)--4 au, while the phase angle varied from about \(12^\circ\) to \(17^\circ\). In contrast, the position angle of the projected Sun–A3 vector decreased from \(313^\circ\) to \(233^\circ\), whereas that of the projected negative heliocentric velocity vector changed from \(101^\circ\) to \(72^\circ\).

\section{Results}
\label{Results}
\subsection{Photometry}
\label{photometry}
To quantify the post-perihelion brightness evolution of comet A3 and to derive the photometric measurements used in the subsequent analysis, we performed aperture photometry on archival images. The flux was measured in a circular aperture of fixed projected linear radius centered on the coma optocenter. This choice allowed us to sample a comparable physical region of the coma at all epochs and to reduce variations caused by the changing geocentric distance. We adopted a projected aperture radius of 40,000 km, which was large enough to include the bright inner coma while limiting contamination from nearby field stars. Over the observing interval, this corresponded to an angular radius of 15.92--16.27 arcsec, depending on the geocentric distance. The sky background was estimated from an annulus with inner and outer projected radii of 60{,}000 and 80{,}000 km, respectively, which lay well outside the source aperture and thus reduced coma contamination while remaining close enough to provide a representative measurement of the local background. Before estimating the background, field stars and cosmic-ray hits within the annulus were identified and masked, after which the sky level was defined as the median of the remaining valid pixel values and subtracted from each image individually.

\begin{figure}
\centering
        \includegraphics[width=1\columnwidth]{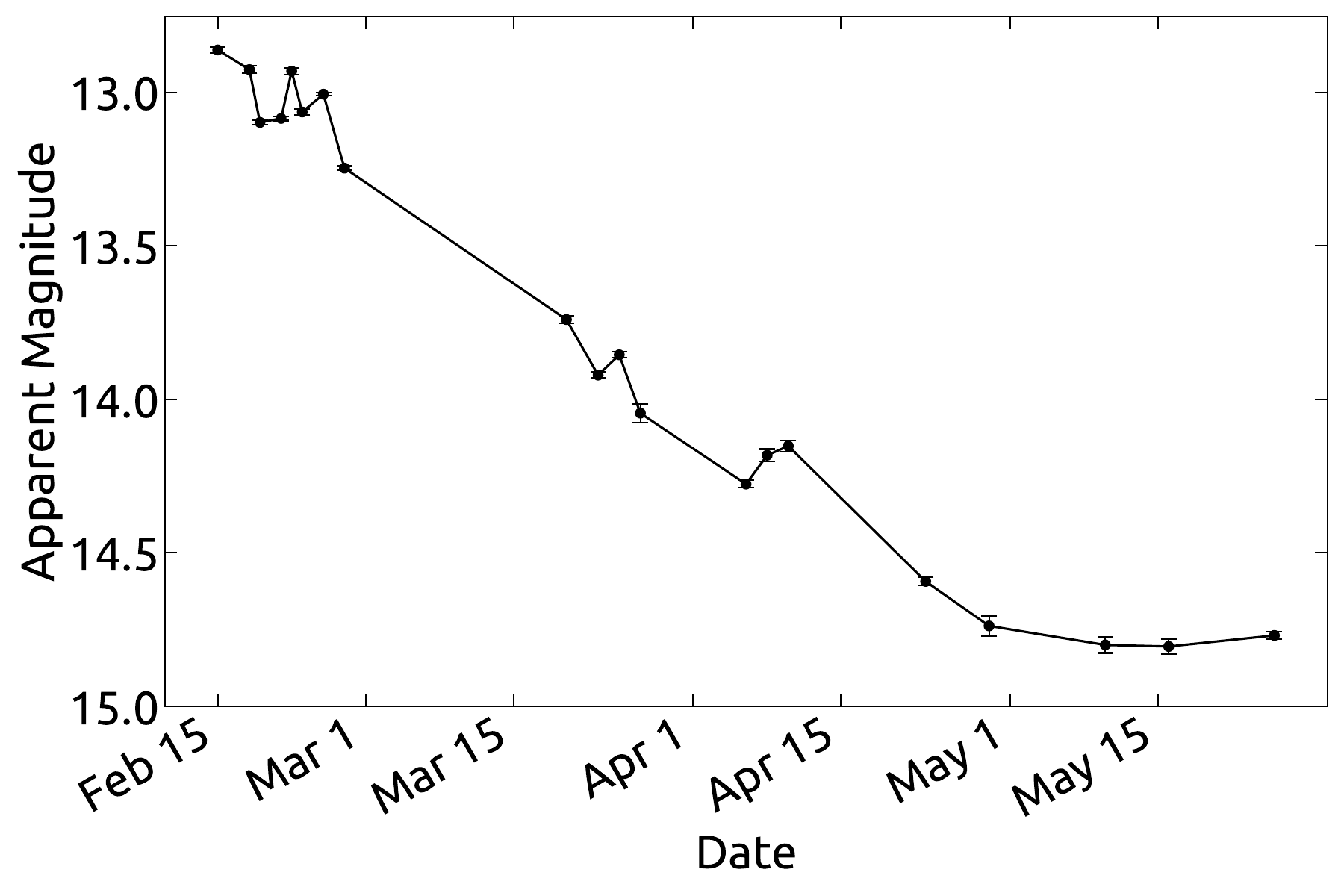}
    \caption{Light curve of comet A3. The horizontal axis represents the observation date, while the vertical axis represents the apparent $r$-band magnitude. Two brightening events are visible around 22 February and 10 April 2025.}
    \label{lightcurve}
\end{figure}

The background-subtracted net counts were converted into apparent \(r\)-band magnitudes, \(m_r\), using the photometric zeropoint recorded in the image header. Table~\ref{tab:photometry_dust} and Fig.~\ref{lightcurve} summarize the resulting photometry and light curve over the interval from 15 February to 26 May 2025. During this period, comet A3 faded from \(m_r = 12.86 \pm 0.01\) to \(14.77 \pm 0.01\) mag as its heliocentric distance increased from 2.64 to 3.90 au. The faintest measurement was obtained on 16 May 2025. Superimposed on the overall post-perihelion fading trend, two short-lived brightenings were present in late February and early April, which reflect a combination of intrinsic activity variations and changes in the observing geometry. The magnitude uncertainties were estimated by propagating the Poisson errors associated with the source and background counts, yielding typical uncertainties of about 0.01--0.03 mag.
\begin{figure*}[!t]
	\sidecaption
	\begin{tabular}{@{}c@{\hspace{0.2cm}}c@{}}
		\includegraphics[width=5.9cm]
		{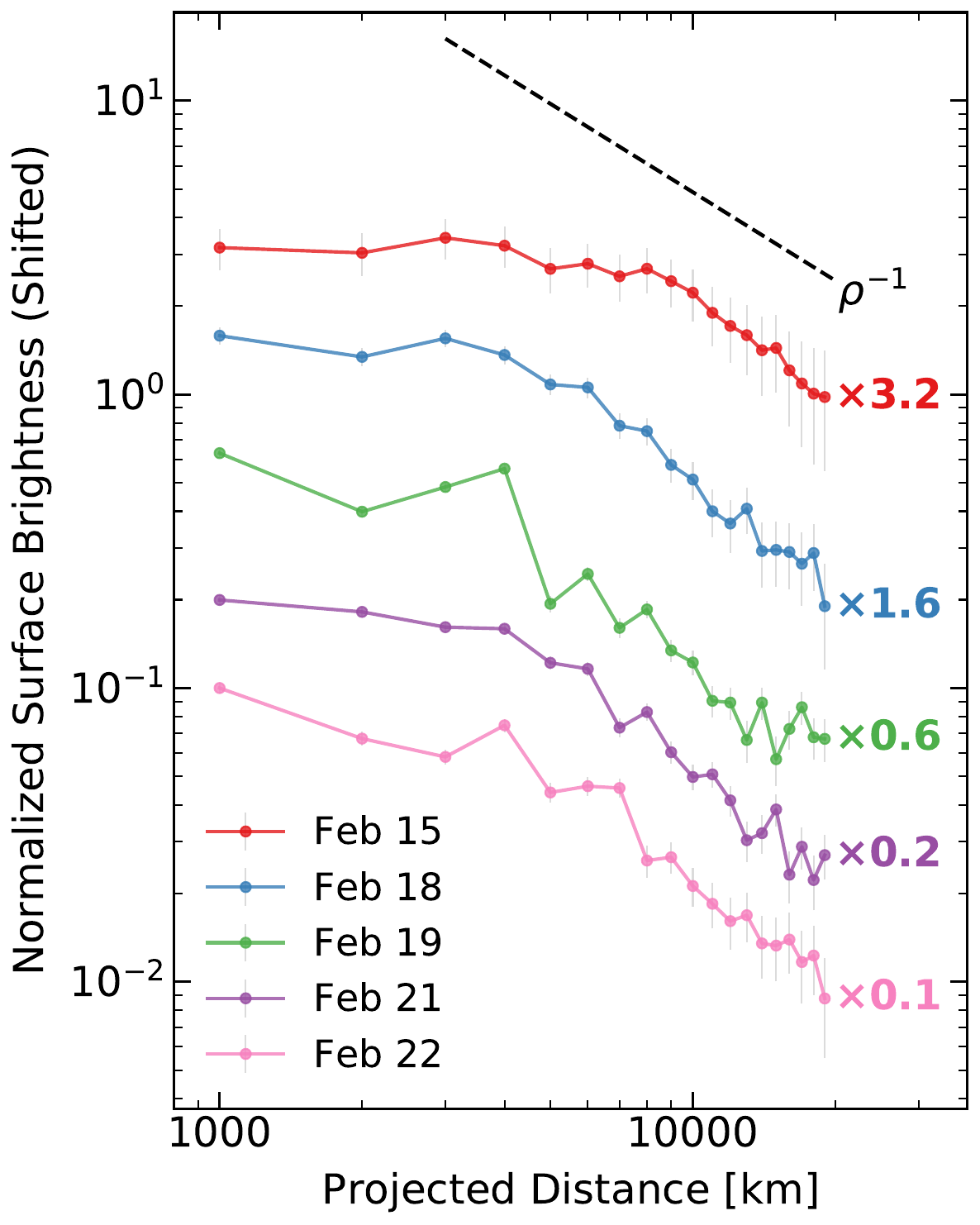}
		&
		\includegraphics[width=5.9cm]
		{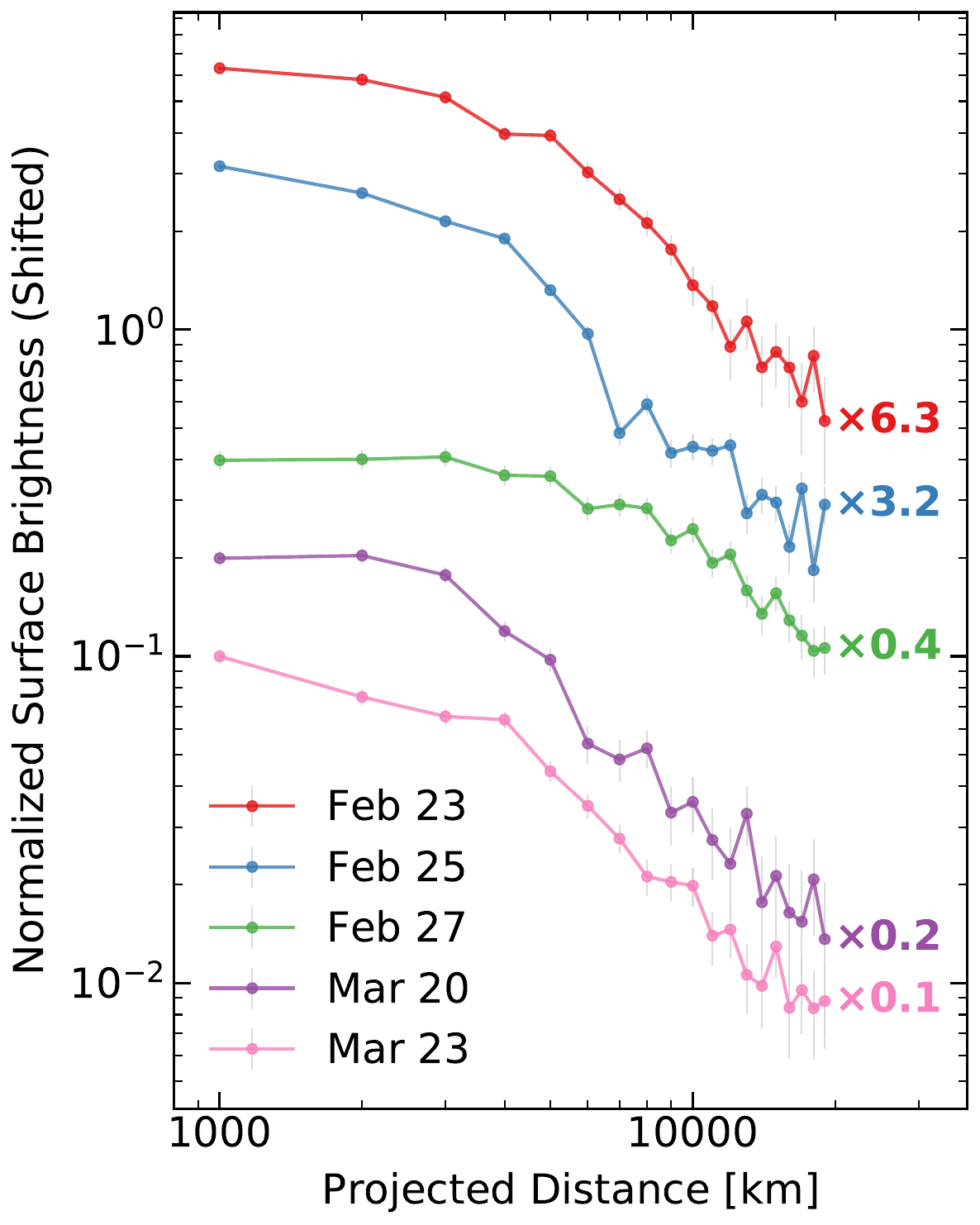}
		\\[0.1cm]
		\includegraphics[width=5.9cm]
		{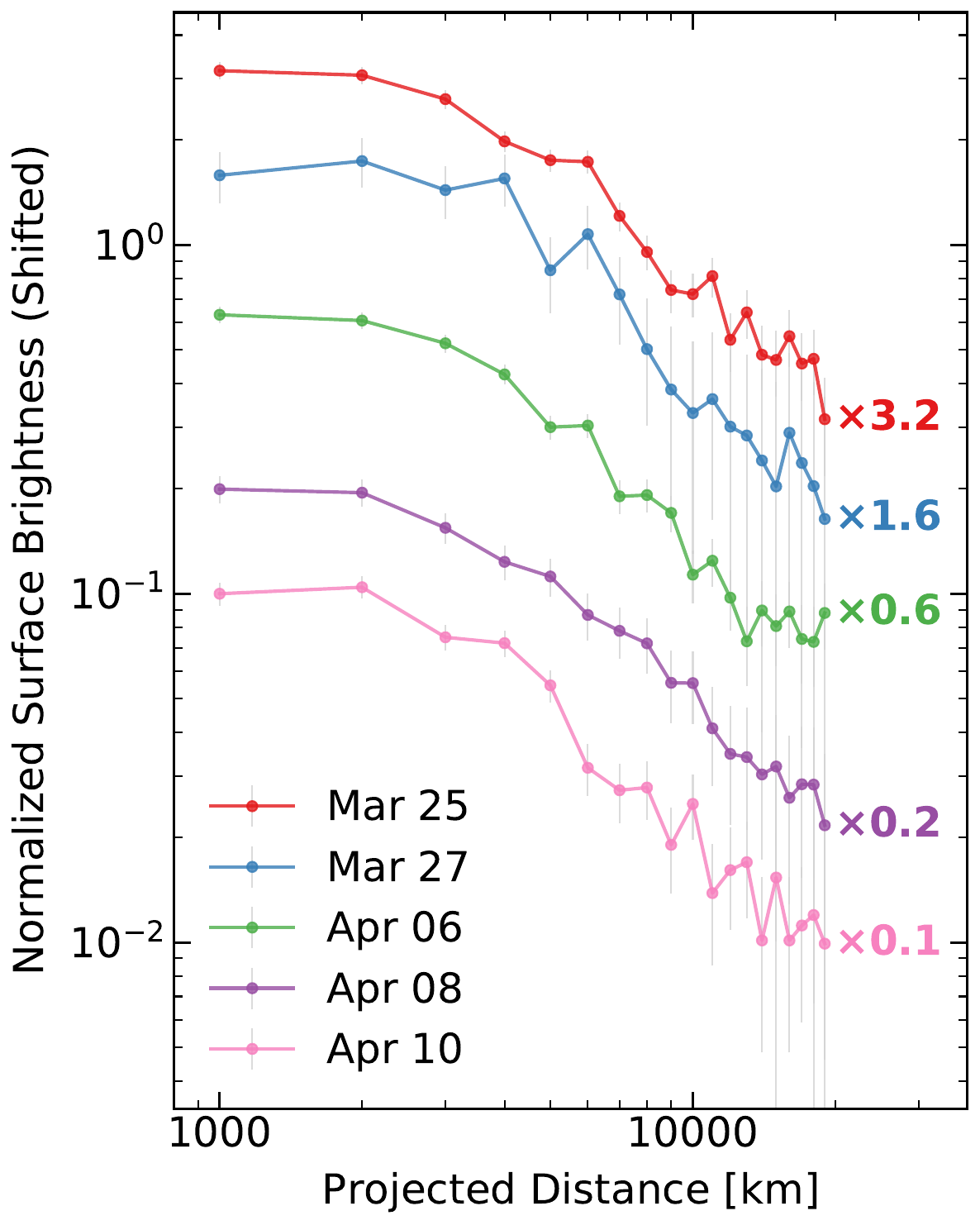}
		&
		\includegraphics[width=5.9cm]
		{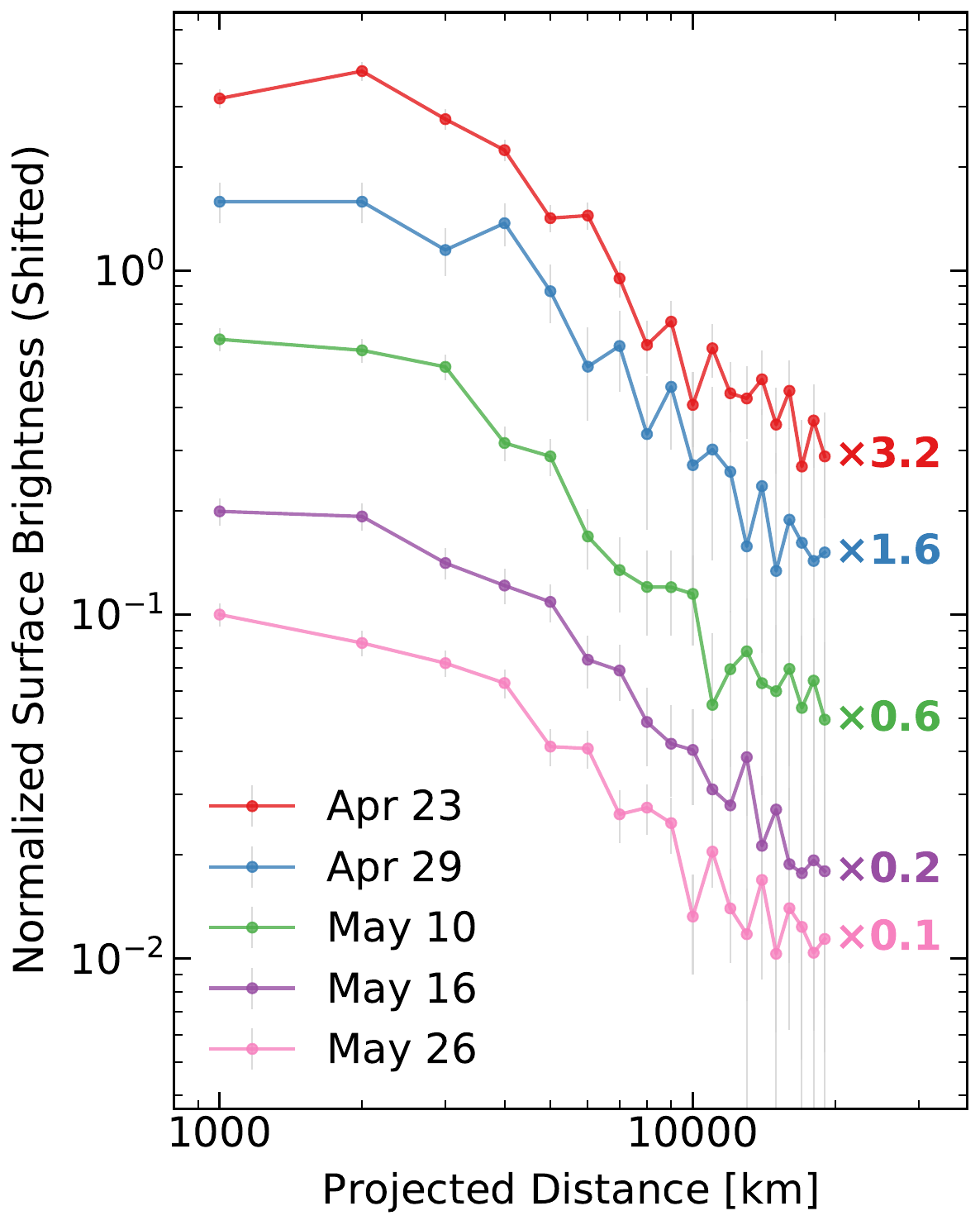}
	\end{tabular}
	\caption{Normalized radial surface brightness profiles of comet A3 as a
		function of projected distance from the nucleus. The profiles are vertically
		offset for clarity, with the offset factors indicated on the right-hand side.
		Error bars represent the corresponding uncertainties. The black dashed segment
		in the upper-right corner of the upper-left panel indicates an arbitrarily
		normalized $I(\rho)\propto\rho^{-1}$ reference slope for a canonical
		steady-state coma.}
	\label{ComaSur}
\end{figure*}

\subsection{Surface brightness}
\label{morphology}
To characterize the radial structure of the post-perihelion coma, we extracted surface-brightness profiles using annular photometry. The azimuthally averaged profiles were derived from 19 concentric annuli centered on the optocenter, each with a projected radial width of 1000~km. Within each annulus, we used the median surface brightness to reduce the influence of contaminating field stars and residual cosmic-ray events. The sky background was estimated from an annulus spanning projected cometocentric radii of 60{,}000--80{,}000~km, after which the resulting profiles were normalized to facilitate comparison among different epochs. Given the seeing-limited resolution, the radial profiles are interpreted in terms of broad radial trends. We also repeated the profile extraction using 3000 km annuli as a robustness check. The resulting profiles preserve the same overall radial behavior and yield slopes that remain shallower than the canonical 1/$\rho$ profile.

\begin{table}
\centering
\renewcommand{\arraystretch}{1.2}
\caption{Measurements of surface brightness profiles for comet A3.}
\label{tab:slope_halo_radius}
\setlength{\tabcolsep}{16pt}
\small
\begin{tabular}{lcc}
\toprule
Date & Slope & Halo radius \\
(UT) & & (km) \\
\midrule
15 February 2025 & -0.31 & $9230 \pm 520$ \\
18 February 2025 & -0.53 & $8010 \pm 470$ \\
19 February 2025 & -0.71 & $7120 \pm 430$ \\
21 February 2025 & -0.62 & $7540 \pm 450$ \\
22 February 2025 & -0.58 & $8380 \pm 500$ \\
23 February 2025 & -0.62 & $9150 \pm 540$ \\
25 February 2025 & -0.84 & $9320 \pm 560$ \\
27 February 2025 & -0.40 & $9170 \pm 620$ \\
20 March 2025 & -0.76 & $8160 \pm 480$ \\
23 March 2025 & -0.53 & $8080 \pm 460$ \\
25 March 2025 & -0.67 & $8190 \pm 490$ \\
27 March 2025 & -0.58 & $8370 \pm 510$ \\
6 April 2025 & -0.71 & $8190 \pm 480$ \\
8 April 2025 & -0.63 & $8610 \pm 520$ \\
10 April 2025 & -0.69 & $9120 \pm 550$ \\
23 April 2025 & -0.80 & $8760 \pm 530$ \\
29 April 2025 & -0.63 & $8900 \pm 570$ \\
10 May 2025 & -0.76 & $9050 \pm 600$ \\
16 May 2025 & -0.68 & $9110 \pm 610$ \\
26 May 2025 & -0.72 & $9330 \pm 650$ \\
\bottomrule
\end{tabular}
\end{table}

Figure~\ref{ComaSur} shows the normalized surface-brightness profiles as a function of projected distance from the nucleus for the 20 epochs between 15 February and 26 May 2025, arranged into four panels with five dates in each. For clarity, the individual profiles in each panel were shifted vertically to avoid overlap while preserving their relative shapes. For comparison, the upper-left panel includes a black dashed \(I(\rho)\propto\rho^{-1}\) reference slope for a canonical steady-state coma. In log--log space, the profiles generally exhibit similar slopes, consistent with an approximately power-law decline of the surface brightness with projected distance, as expected primarily from geometric dilution in the coma \citep{jewitt1987surface}. The associated uncertainties are indicated by error bars. Each profile was fitted with a power-law relation of the form \(S(\rho)\propto \rho^{m}\) using a least-squares method, and the resulting indices \(m\) are listed in Table~\ref{tab:slope_halo_radius}. The derived slopes are consistently shallower than the canonical value of $m=-1$ expected for a steady-state coma in free expansion. These deviations indicate that the post-perihelion coma cannot be described by a simple steady outflow with constant dust properties. A detailed discussion of the possible physical causes is presented in Sect.~\ref{discussion}.

\section{Discussion}
\label{discussion}
\subsection{Dust mass loss}
\label{Dust properties}
Having established the light curve in Sect.~\ref{photometry}, we converted the observed \(r\)-band magnitudes to absolute magnitudes in order to reduce the effects of changing observing geometry and better trace the intrinsic post-perihelion evolution of comet A3. The conversion is given by
\begin{equation}
H_r = m_r - 5\log_{10}(r_{\rm h}\Delta) - \beta \alpha.
\label{eq:abs_mag}
\end{equation}
Here, \(r_{\rm h}\) and \(\Delta\) correspond to the distances from the Sun and Earth in au, \(\alpha\) is the phase angle in degrees, and \(\beta = 0.028 \pm 0.004~\mathrm{mag\,deg^{-1}}\) is the linear phase coefficient adopted from \citet{liu2025great}. As listed in Table~\ref{tab:photometry_dust}, \(H_r\) increased from \(7.75 \pm 0.05\) mag on 15 February 2025 to \(8.72 \pm 0.06\) mag on 26 May 2025, indicating a decline in the geometry-corrected brightness of the dust coma during the post-perihelion interval.

We also estimated the contribution of the nucleus to the aperture photometry. Adopting \(R_{\rm N}=5.9\) km estimated by \citet{liu2025great}, \(p_r=0.1\), and \(\beta=0.028~{\rm mag~deg^{-1}}\), we obtained a projected nucleus cross-section of \(C_{\rm N} = 109.4~{\rm km^{2}}\), corresponding to \(H_{r,\rm N}=12.62\) mag. Comparison with the total scattering cross-sections in Table~\ref{tab:photometry_dust} shows that the nucleus contributes approximately \(1.1\%{-}3.0\%\) of the total flux measured within the \(40{,}000\) km aperture. Subtracting this contribution would make the coma magnitudes fainter by only \(0.012{-}0.033\) mag. The nucleus contribution to the integrated photometry is therefore minor and does not materially affect the derived temporal trends. Given this small contribution of the nucleus, we treat the measured aperture brightness as dust dominated. To quantify the evolution of the dust coma, we estimated the effective scattering cross-section, \(C_{\rm e}\) (in km$^2$), from the absolute magnitude as
\begin{equation}
C_{\rm e} = \frac{2.24 \times 10^{16}\pi}{p_r} \, 10^{0.4(m_{\odot,r} - H_r)} ,
\label{eq:cross_section}
\end{equation}
where we adopted a geometric albedo of \(p_r = 0.1\) \citep{jewitt2019initial} and a solar \(r\)-band magnitude of \(m_{\odot,r} = -26.9\) mag \citep{willmer2018absolute}. Over the observing interval, \(C_{\rm e}\) declined from about \(9.68\times10^3\) km$^2$ to \( 3.99\times10^3\) km$^2$, indicating a substantial reduction in the scattering area contained within the fixed aperture as comet A3 receded from the Sun from \(r_{\rm h}=2.64\) to 3.90 au. Superimposed on this secular decline, short-lived enhancements were identified around 22 February and 10 April 2025. Relative to the pre-enhancement levels, the inferred dust cross-section increased by approximately 22\% and 15\%, respectively, corresponding to brightenings of 0.17--0.22 mag in \(H_r\) (Table~\ref{tab:photometry_dust}).

\begin{figure}
\centering
        \includegraphics[width=1\columnwidth]{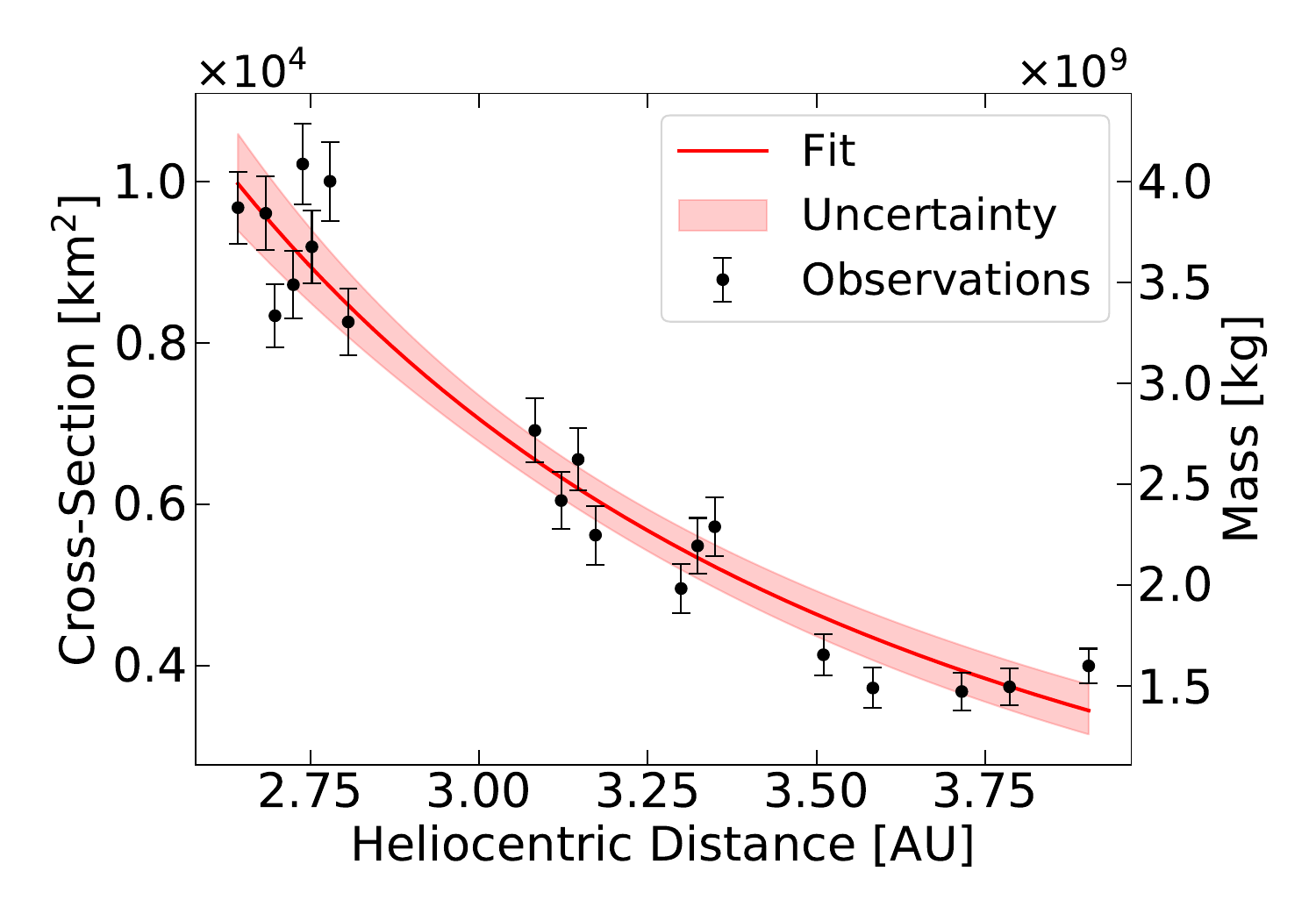}
    \caption{Effective cross-section of comet A3 as a function of heliocentric distance. The black points with error bars show the measurements derived from the photometry, the red curve shows the best-fit relation, and the shaded region indicates the corresponding confidence band. The right-hand axis gives the equivalent dust mass.}
    \label{C_e_curve_with_mass}
\end{figure}

To describe the evolutionary trend of dust activity with respect to the heliocentric distance, we fitted the measurements with a power law of the form
\begin{equation}
C_{\rm e}(r_{\rm h}) = C_0\, r_{\rm h}^{-s},
\label{eq:powerlaw}
\end{equation}
where $C_0$ is the formal normalization at $r_{\rm h}=1$ au and $s$ describes the radial decline of the scattering cross-section. The resulting best-fit parameters are $C_0 = (2.1 \pm 0.7)\times10^{5}\ \mathrm{km^2}$ and $s = 3.1 \pm 0.3$, respectively. The fitted relation and its $2\sigma$ confidence band are shown in Fig.~\ref{C_e_curve_with_mass}.

\begin{figure}
\centering
        \includegraphics[width=1\columnwidth]{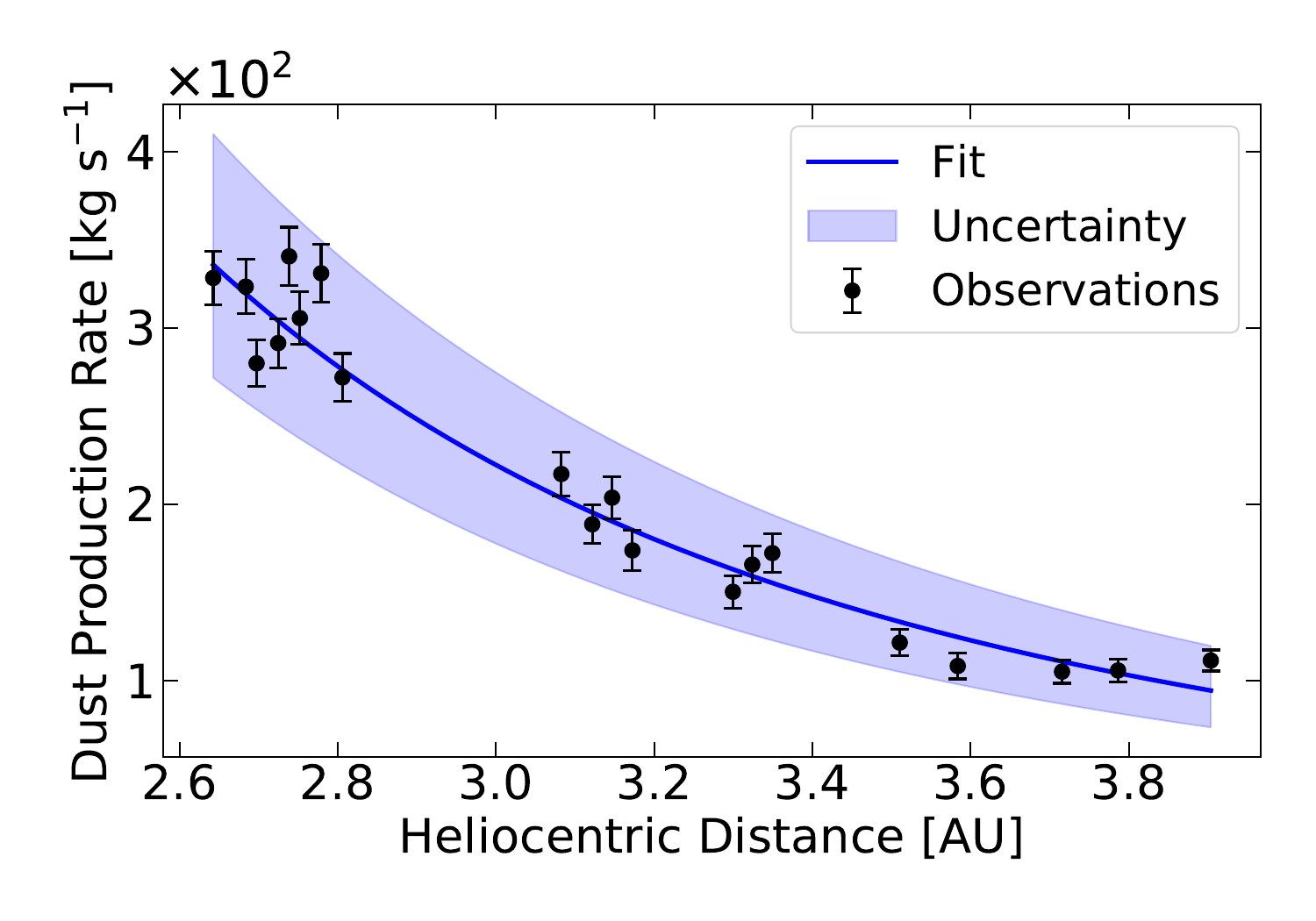}
    \caption{Dust production rate of comet A3 as a function of heliocentric distance. The black points with error bars show the measurements derived from the photometry, the blue curve shows the best-fit relation, and the shaded region indicates the corresponding confidence band. }
    \label{C_e_curve_with_mass_RATE}
\end{figure}

Figure~\ref{C_e_curve_with_mass} also shows the equivalent dust mass corresponding to the measured effective scattering cross-section. Following \citet{jewitt2009active}, we assumed a differential power-law grain-size distribution, $dN/da \propto a^{-q}$, and calculated the dust mass as
\begin{equation}
M_{\rm d}
=
\frac{4}{3}\rho_{\rm d} C_{\rm e}
\frac{
	\displaystyle \int_{a_{\min}}^{a_{\max}} a^{3-q}\, da
}{
	\displaystyle \int_{a_{\min}}^{a_{\max}} a^{2-q}\, da
}.
\label{eq:mass}
\end{equation}
Following \citet{liu2025great}, we adopted $a_{\min}=20\,\mu{\rm m}$, $a_{\max}=10\,{\rm mm}$, and $q=3.4$, together with a dust bulk density of $\rho_{\rm d}=500\,{\rm kg\,m^{-3}}$. The mass estimate thus incorporates the inferred grain-size distribution rather than treating the dust as a monodisperse population. However, because the post-perihelion grain-size distribution is not independently constrained by our observations, the inferred dust mass is sensitive to the adopted lower and upper size limits. To further quantify the strength of the cometary activity, we estimated the dust mass production rate, \(\dot{M}_{\rm d}\), from the dust mass contained within the photometric aperture. Assuming steady radial transport of dust through the aperture, the production rate was calculated as
\begin{equation}
\dot{M}_{\rm d} = \frac{M_{\rm d}}{\tau_r},
\label{eq:dust_production}
\end{equation}
where \(\tau_r\) is the residence time of dust grains within the aperture. We adopted \(\tau_r = L / v_{\rm ej}\), where \(L = 4\times10^4\) km is the projected aperture radius and \(v_{\rm ej}\) is the dust ejection velocity. Following \citet{liu2025great}, the latter was assumed to scale with heliocentric distance as
\begin{equation}
v_{\rm ej} = v~(a_{\mu \rm m}) \left( \frac{r_{\rm h}}{r_{\rm h,ref}} \right)^{-1/2},
\end{equation}
with \(v~(a_{\mu \rm m}) = 88~a_{\mu \rm m}^{-1/2}~\mathrm{m\,s^{-1}}\) at \(r_{\rm h,ref} = 2.36\) au. For the velocity calculation, we adopted an equivalent mean grain radius of
	$\bar{a}\sim600~\mu{\rm m}$ (i.e., $a_{\mu{\rm m}}\sim600$), derived from the same
	grain-size distribution used in Eq.~\ref{eq:mass}. Under these assumptions, the derived dust mass production rate decreased with increasing heliocentric distance, indicating a gradual weakening of the post-perihelion dust activity. To characterize the global trend, we fitted the measurements with a weighted power-law model of the form
\begin{equation}
\dot{M}_{\rm d} = A r_{\rm h}^{-n},
\label{eq:production_fit}
\end{equation}
where the weights were defined from the propagated uncertainties in \(C_{\rm e}\) alone. The best-fit parameters are $A = (7.91 \pm 0.87)\times10^3~\mathrm{kg\,s^{-1}}$ and $n = 3.41 \pm 0.10$. The derived production rates, together with the best-fit relation and its confidence band, are shown in Fig.~\ref{C_e_curve_with_mass_RATE}. This dependence is significantly steeper than the simple \(r_{\rm h}^{-2}\) scaling expected if the activity is driven by volatile species such as CO$_2$, and is closer to the weaker end of the $r_{\rm h}^{-3.4}$ to $r_{\rm h}^{-14.5}$ range expected for H$_2$O sublimation over comparable heliocentric distances \citep{jewitt2025preperihelion}. Extrapolating this relation back to the perihelion epoch and integrating it up to the last observing epoch yields a cumulative dust mass loss of
\begin{equation}
M_{\rm loss} = \int_{t_1}^{t_2} \dot{M}_{\rm d}(t)\,dt,
\end{equation}
where \(t_1\) and \(t_2\) are the perihelion epoch (27 September 2024) and the last observing epoch (26 May 2025), respectively. In this way, we obtained \(M_{\rm loss}=3.0\times10^{11}\) kg for the post-perihelion phase. This value should be interpreted as an order-of-magnitude estimate. The absolute mass scale depends on the adopted grain-size distribution, albedo, bulk density, activity intensity and velocity law.

For comparison, this mass corresponds to an equivalent globally averaged erosion depth of
\begin{equation}
\Delta R = \frac{M_{\rm loss}}{4\pi R_{\rm N}^2 \rho_{\rm N}},
\end{equation}
where \(R_{\rm N}\) is the nucleus radius and \(\rho_{\rm N}\) is the bulk density of the nucleus. Adopting \(R_{\rm N}=5.9\) km and \(\rho_{\rm N}=500~\mathrm{kg\,m^{-3}}\) gives \(\Delta R \approx 1.4\) m. Although this level of erosion is modest when averaged over the entire nucleus, the local erosion depth could have been substantially larger if the activity was confined to discrete active regions. For example, if sublimation was restricted to an active fraction of order 10\% of the surface, the corresponding local erosion depth would increase to \(\sim 14\) m. Such localized excavation could help remove a superficial devolatilized layer, expose deeper volatile-rich material, and thereby facilitate localized outgassing and the release of ice-bearing grains into the coma.

\subsection{Shallow surface-brightness profiles}
\label{Tail Dynamical Modelling}
The logarithmic slopes derived from the radial surface-brightness profiles deviate from the canonical steady-state expectation. As listed in Table~\ref{tab:slope_halo_radius}, all measured slopes are shallower than $m=-1$, the value expected for a steady coma expanding at constant velocity with unchanged scattering properties. Notably, comparable profile flattening was not detected during the pre-perihelion leg analyzed by \citet{liu2025great}, indicating that the shallow profiles identified here are characteristic of the post-perihelion observations. One possible explanation is delayed fragmentation of dust aggregates after their release from the nucleus. Several mechanisms may produce such delayed fragmentation. In ice-bearing aggregates, inefficient gas escape can cause sublimation pressure to build within the particle. If this pressure exceeds the aggregate's mechanical strength, the parent grain can disrupt into smaller fragments \citep{markkanen2020thermophysical}. Sublimation may also weaken intergrain contacts, making the aggregate more susceptible to subsequent breakup \citep{schuckart2025release}. Irregular, weak aggregates may additionally undergo rotational disruption after being spun up by radiative torques \citep{herranen2020rotational}. Rapid heating, decompression, phase transitions, and electrostatic stresses may provide further fragmentation pathways \citep{clark2004release}.

For comet A3, mechanisms related to ice sublimation are relevant because of the candidate icy-grain halo discussed below. As parent aggregates move outward, they may break into smaller and more numerous daughter particles, increasing the total effective scattering cross-section of the outflow and flattening the projected brightness profile. To illustrate qualitatively how this process can modify the coma profile, we consider an extension of the steady-state framework of \citet{jewitt1987surface}.

In an optically thin coma, the observed surface brightness is the line-of-sight projection of the three-dimensional scattering emissivity. For a steady outflow with constant grain velocity, mass conservation implies \(\varrho_{\rm d}(r)\propto r^{-2}\), where \(r\) is the true cometocentric distance. If fragmentation modifies the effective scattering cross-section per unit dust mass along the flow, the emissivity can be written as \(j(r)\propto \varrho_{\rm d}(r)\kappa_{\rm sca}(r)\), where $\kappa_{\rm sca}(r)$ is the mass scattering opacity. The projected surface-brightness profile is then
\begin{equation}
I(p)\propto \int_{-\infty}^{+\infty}\frac{\kappa_{\rm sca}\!\left(\sqrt{p^{2}+l^{2}}\right)}{p^{2}+l^{2}}\,dl,
\end{equation}
where $p$ is the projected cometocentric distance and $l$ is the line-of-sight coordinate. In the canonical case of constant $\kappa_{\rm sca}$, this reduces to $I(p)\propto p^{-1}$.

To relate this explicitly to fragmentation, we define \(N_{\rm f}(r)\) as the cumulative mean number of daughter fragments produced from a single parent grain by the time it has reached distance \(r\). In the equal-mass fragmentation limit, mass conservation gives \(a_{\rm d}=a_{\rm p}N_{\rm f}^{-1/3}\), where \(a_{\rm p}\) and \(a_{\rm d}\) are the parent- and daughter-grain radii. Assuming that the daughter grains remain in the Mie-like scattering regime characterized by size parameter $x = 2\pi a_{\mathrm{d}}/\lambda \gtrsim 1$, where $\lambda$ is the
observing wavelength, the scattering cross-section of an individual fragment scales as $\sigma_{\rm sca}\propto a_{\rm d}^{2}$. The total scattering cross-section of all fragments therefore scales as $N_{\rm f}a_{\rm d}^{2}\propto a_{\rm p}^{2}N_{\rm f}^{1/3}$. Since the total dust mass is conserved, the mass scattering opacity increases as fragmentation proceeds, giving $\kappa_{\rm sca}(r)\propto N_{\rm f}(r)^{1/3}$. In this case, additional scattered light is contributed at larger cometocentric distances, and the logarithmic slope $m(p)= {\rm d}\ln I/{\rm d}\ln p$ becomes less negative than the canonical value $-1$, meaning the surface-brightness profile flattens. Fragmentation therefore provides a plausible explanation for the shallower-than-canonical profiles measured in comet A3. 

However, if fragmentation produces daughter grains small enough to enter the Rayleigh regime ($x \ll 1$), the scattering cross-section declines steeply as $\sigma_{\rm sca} \propto a_{\rm d}^{6}/\lambda^{4}$. In this limit, the mass scattering opacity decreases as fragmentation proceeds, with
\begin{equation}
\kappa_{\rm sca}(r) \propto N_{\rm f}(r)^{-1}.
\end{equation}
This behavior would have the opposite effect to the Mie-like case and would tend to steepen the surface-brightness profile. In addition, smaller fragments are more strongly affected by radiation pressure, which can lead to size-dependent velocity evolution and departures from the assumed $r^{-2}$ density profile. These effects complicate the interpretation of the measured slopes. A fully self-consistent treatment that couples grain fragmentation, size-dependent dust dynamics, and scattering properties would be required to model the coma brightness profile quantitatively. The analytic framework used here should therefore be regarded as a first-order approximation.

\begin{figure}
\centering
        \includegraphics[width=1\columnwidth]{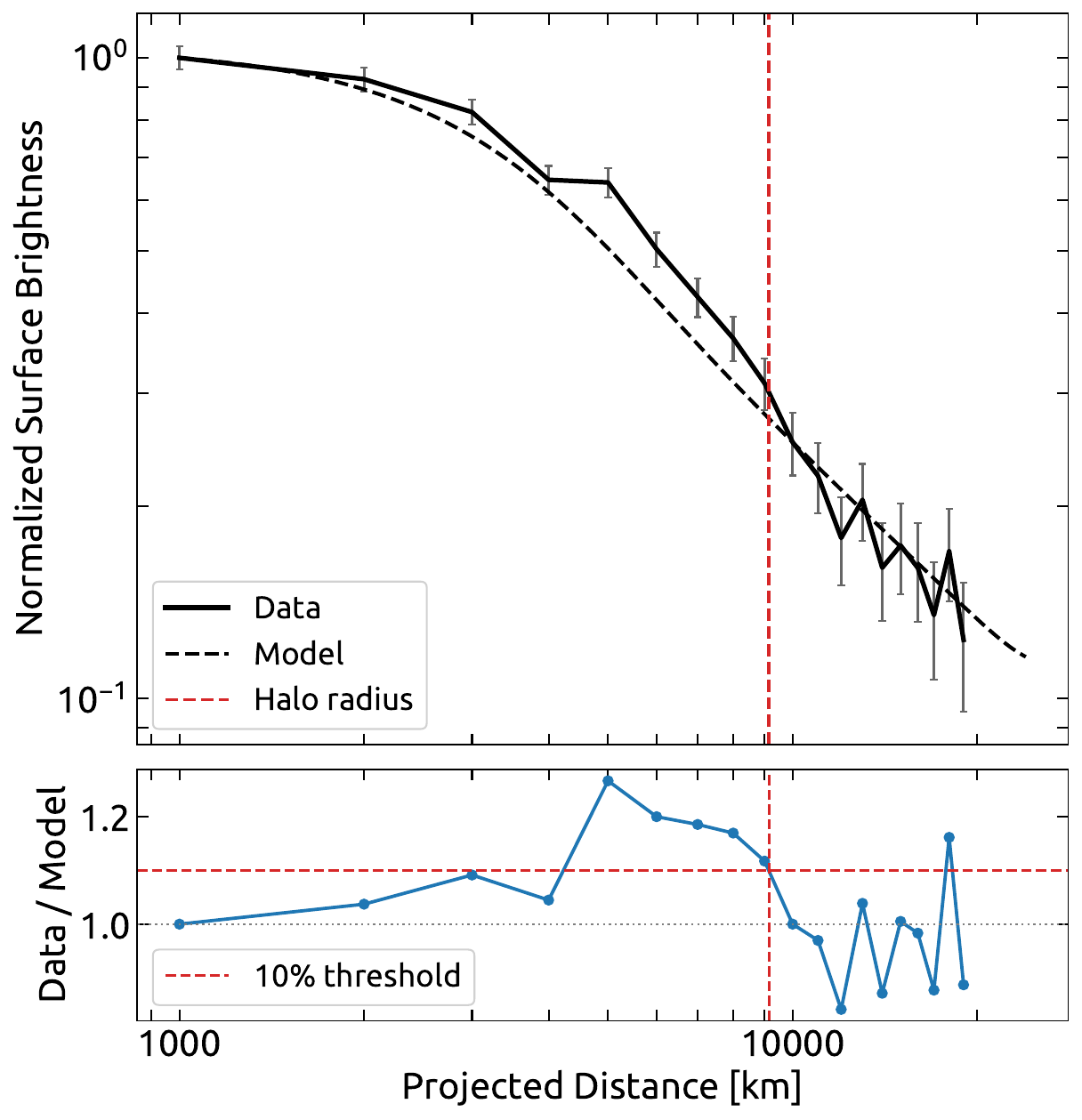}
    \caption{Comparison between the observed and modeled radial surface-brightness
    	profiles on 23 March 2025. Upper panel: observed normalized surface-brightness profile (solid black line with error bars) compared with the seeing-convolved model consisting of a smooth coma and an unresolved central point-source component (black dashed line). The red vertical dashed line marks the inferred halo radius. Lower panel: ratio between the observed and modeled profiles as a function of projected cometocentric distance. The horizontal red dashed line indicates the 10\% threshold used to define the halo radius, namely the first distance at which the observed profile falls to within 10\% of the model prediction.}
    \label{modelsu}
\end{figure}

\begin{figure}
\centering
        \includegraphics[width=1\columnwidth]{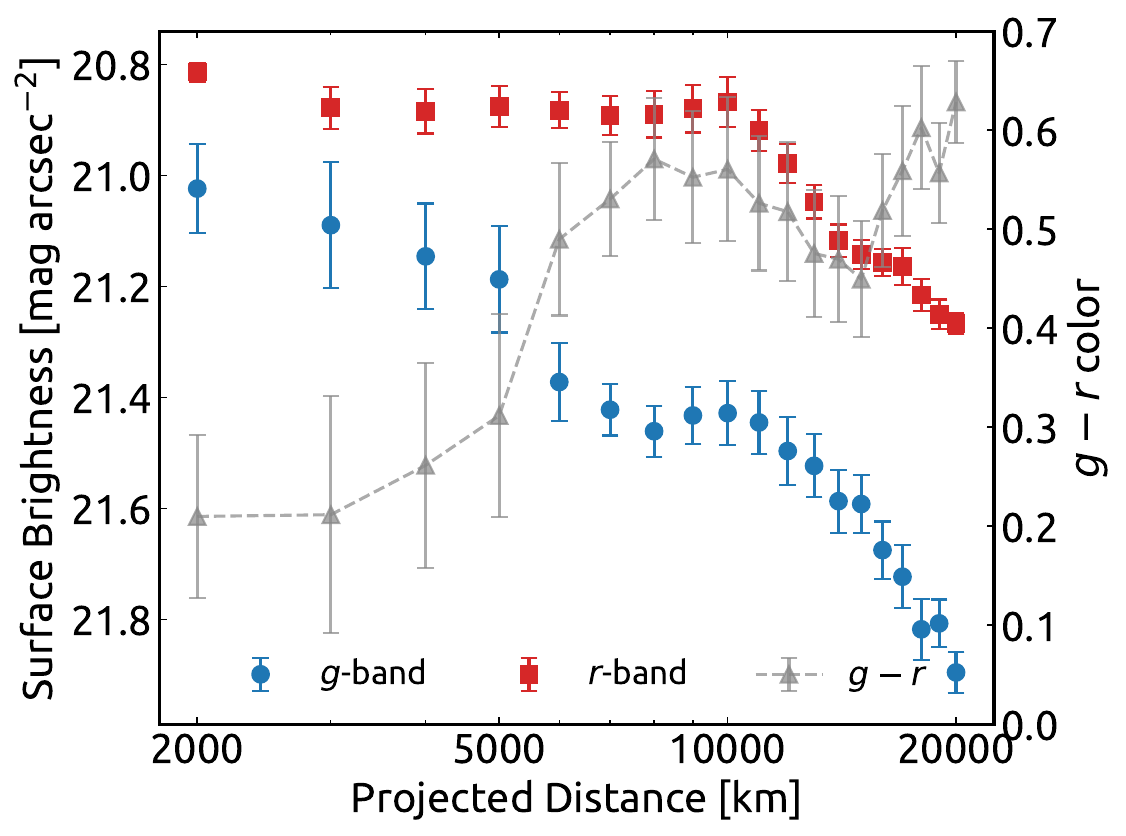}
    \caption{Radial surface-brightness profiles in the \(g\) and \(r\) bands, together with the corresponding \(g-r\) color profile. Distances are projected cometocentric distances.}
    \label{color}
\end{figure}

\subsection{Icy-grain halo}
\label{sec:halo}
The radial surface-brightness profiles shown in Fig.~\ref{ComaSur} exhibit a persistent central excess. Because this feature is present at all epochs, it is unlikely to result from random fluctuations or reduction artifacts and instead appears to be an intrinsic property of the inner coma. Such an excess could partly arise from an unresolved central point-source component, whose flux is redistributed into the innermost coma bins by atmospheric seeing, or it could indicate an additional extended component near the nucleus. To test these possibilities, we constructed a forward model consisting of a smooth coma with a power-law surface-brightness profile and an unresolved central point-source component. Both components were convolved with an effective Gaussian PSF whose FWHM was measured from nearby field stars in each image. Here, the point-source component represents flux on spatial scales smaller than the seeing disk and may include contributions from both the solid nucleus and the innermost unresolved coma. Explicitly, the seeing-convolved model was written as
\begin{equation}
S_{\rm mod}(\rho)
=
A_{\rm c}
\left[
\left(
\frac{\rho}{\rho_{0}}
\right)^{m}
\otimes P_{\rm PSF}
\right](\rho)
+
A_{\rm p}P_{\rm PSF}(\rho),
\label{eq:coma_point_model}
\end{equation}
where $\rho$ is the projected cometocentric distance, $\rho_{0}=10,000$~km is the reference distance, and $P_{\rm PSF}(\rho)$ is the effective Gaussian PSF. The first term on the right-hand side represents the smooth coma, described intrinsically by a power-law profile with logarithmic slope $m$ and convolved with the PSF. The second term represents the unresolved central point-source component, whose observed profile follows the PSF. The quantities $A_{\rm c}$ and $A_{\rm p}$ are the normalizations of the smooth-coma and point-like components, respectively, and $\otimes$ denotes convolution. In the numerical implementation, the power-law profile was regularized only within the central model-grid cell to avoid its formal divergence at $\rho=0$. The outer coma slope was determined by fitting the observed profile over projected cometocentric distances of 10{,}000--19{,}000 km, and the relative weights of the smooth-coma and central point-like components were then jointly determined by requiring the convolved model to match the observed brightness levels at projected distances of 1000 and 10{,}000 km. A radial profile extracted from the convolved model was then compared with the observed one.

Figure~\ref{modelsu} illustrates the procedure for the epoch with the highest signal-to-noise ratio (23 March 2025). At all epochs, the model underpredicted the observed profile, indicating that a smooth coma plus an unresolved central point-source component is insufficient to account for the observed central excess. We verified that changes in the PSF parametrization, optocenter, background annulus, and position-angle sector did not remove the central excess. This excess therefore suggests the presence of an additional near-nucleus component. The color profile provides an independent check on this interpretation. For the color comparison, we used the $g$- and $r$-band images obtained on 23 March 2025. The two exposures were taken on the same night, at 11:04:00 UTC in the $g$ band and 12:23:18 UTC in the $r$ band, with a time separation of 1.32 hr. The seeing FWHM values are $2.43^{\prime\prime}$ and $2.19^{\prime\prime}$ in the $g$ and $r$ bands, respectively. To minimize seeing-induced color gradients, the $r$-band image was convolved with a Gaussian kernel to match the broader $g$-band PSF. The brightness profiles of the two images were extracted with the same optocenter, projected annuli, background annulus, and field-star masks. The $g-r$ color profile was computed from the difference between the corresponding surface-brightness profiles, and its uncertainty was obtained by propagating the independent errors in the two bands as $\sigma_{g-r}=(\sigma_g^2+\sigma_r^2)^{1/2}$. 

Figure~\ref{color} shows that the inner coma is less red than the outer coma, with lower \(g-r\) values at small projected distances. In a	dust-dominated coma, larger and more slowly moving grains tend to remain concentrated near the nucleus, whereas smaller grains can extend to larger cometocentric distances and are associated with bluer colors. The innermost coma is therefore often expected to appear redder than the outer regions \citep{opitom2019first,kwon2023coma}. The opposite radial color trend observed in comet A3 resembles centrally concentrated blue color gradients previously reported for comet Halley and C/2013 A1 (Siding Spring), where they were associated with volatile or icy grains in the inner coma \citep{campins1989infrared,li2014constraining}. Moreover, the color gradient in comet A3 extends over approximately the same spatial scale as the surface-brightness excess. The spatial coincidence of the photometric and color anomalies supports the interpretation of a compact near-nucleus halo of fresh or ice-rich grains, in which sublimating icy particles contribute additional scattering cross-section. Such a component could also act as an extended source of water in the inner coma, providing a possible contribution to the elevated apparent water-production rate reported by \citet{ahuja2024molecular} and discussed by \citet{sekanina2024inevitable}.

To quantify the radial extent of the excess component, we defined the halo radius as the first projected distance at which the ratio of the observed profile to the model profile decreased from values \(>1.10\) to \(\leq 1.10\) (e.g., \citealt{stevenson2012near}). Because the threshold crossing generally occurred between two adjacent annular bins, the halo radius was obtained by linear interpolation between the two neighboring bin centers. This empirical definition is intended only as a practical measure of the radial extent of the excess component, not as a physical boundary. The uncertainties reflect the finite 1000 km radial bin size and the local scatter in the observed-to-model ratio around the point where it crosses the 10\% threshold. The resulting halo radii are listed in Table~\ref{tab:slope_halo_radius}. The halo radius did not vary monotonically with heliocentric distance, showing local increases around the late-February and early-April activity enhancements seen in the light curve. This temporal correspondence suggests that renewed activity may have episodically replenished near-nucleus icy grains and thereby modulated the apparent extent of the halo.

\subsection{Thermal model for icy grain radius loss}
We computed the radius-loss rate of icy grains exposed to solar radiation following the approach described by \citet{moreno2018icy}. The particles were treated as compact spheres of radius $a$ at heliocentric distance $r_{\rm h}$. We considered water-ice grains that are either pure or mixed with refractory inclusions. The inclusions modify the optical absorption of the grains and hence their equilibrium temperatures. For a grain with complex refractive index of $\tilde{m}(\lambda)=n(\lambda)+ik(\lambda)$, the absorption efficiency \(Q_{\rm abs}(\lambda,a)\) was computed using Mie theory. For mixed grains, water ice was treated as the host medium and the refractory components as inclusions. The effective dielectric constant was calculated using the Maxwell-Garnett relation
\begin{equation}
\frac{\epsilon_{\rm eff}-\epsilon_0}
{\epsilon_{\rm eff}+2\epsilon_0}
=
\sum_i f_i
\frac{\epsilon_i-\epsilon_0}
{\epsilon_i+2\epsilon_0},
\end{equation}
where \(\epsilon_0\) is the dielectric constant of water ice, \(\epsilon_i\) and \(f_i\) are the dielectric constant and volume fraction of the \(i\)-th inclusion, and $\tilde{m}_{\rm eff}=\epsilon_{\rm eff}^{1/2}$. The resulting $\tilde{m}_{\rm eff}(\lambda)$ was then used to calculate $Q_{\rm abs}$.

\begin{figure}
\centering
    \includegraphics[width=\columnwidth]{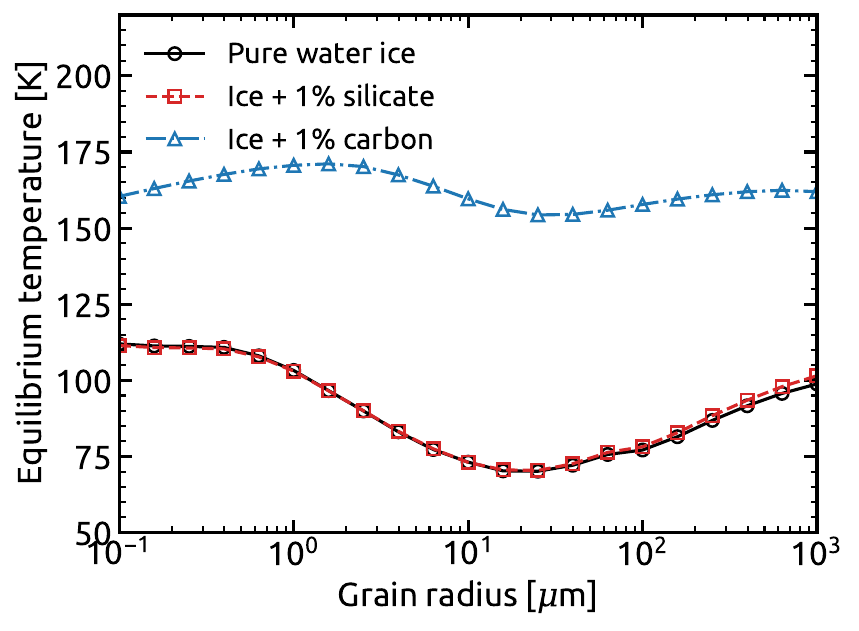}
    \caption{Equilibrium temperature of water-ice grains with different radii at a heliocentric distance of $3\,\mathrm{au}$. The black circles represent pure water ice, the red squares represent ice containing 1\% silicate, and the blue triangles represent ice containing 1\% carbon.}
    \label{temp}
\end{figure}

\begin{figure}
	\centering
	\includegraphics[width=\columnwidth]{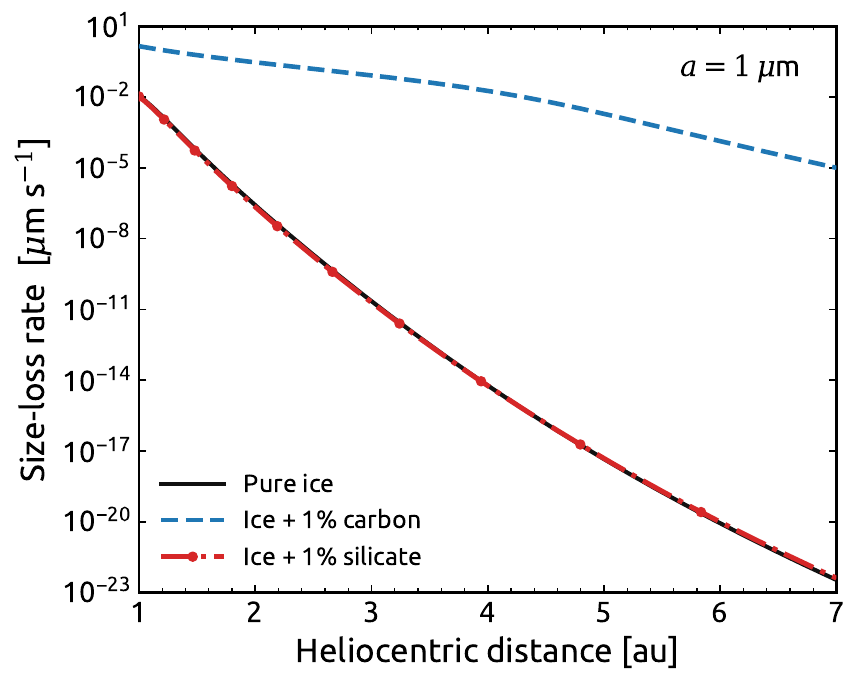}
	\caption{Size-loss rate of water-ice grains with a radius of $a = 1\,\mu\mathrm{m}$ as a function of heliocentric distance. The black solid line represents pure water ice, the blue dashed line represents ice containing 1\% carbon, and the red dash-dotted line represents ice containing 1\% silicate. }
	\label{sizeloss}
\end{figure}
The equilibrium temperature, \(T\), was obtained from the energy balance
\begin{equation}
\frac{1}{r_{\rm h}^2}
\int_0^\infty S_\odot Q_{\rm abs}\,d\lambda
=
4\int_0^\infty Q_{\rm abs}\pi B_\lambda\,d\lambda
+4L_vZ ,
\label{eq:thermal_balance}
\end{equation}

\begin{figure*}
	\centering
	\includegraphics[width=2\columnwidth]{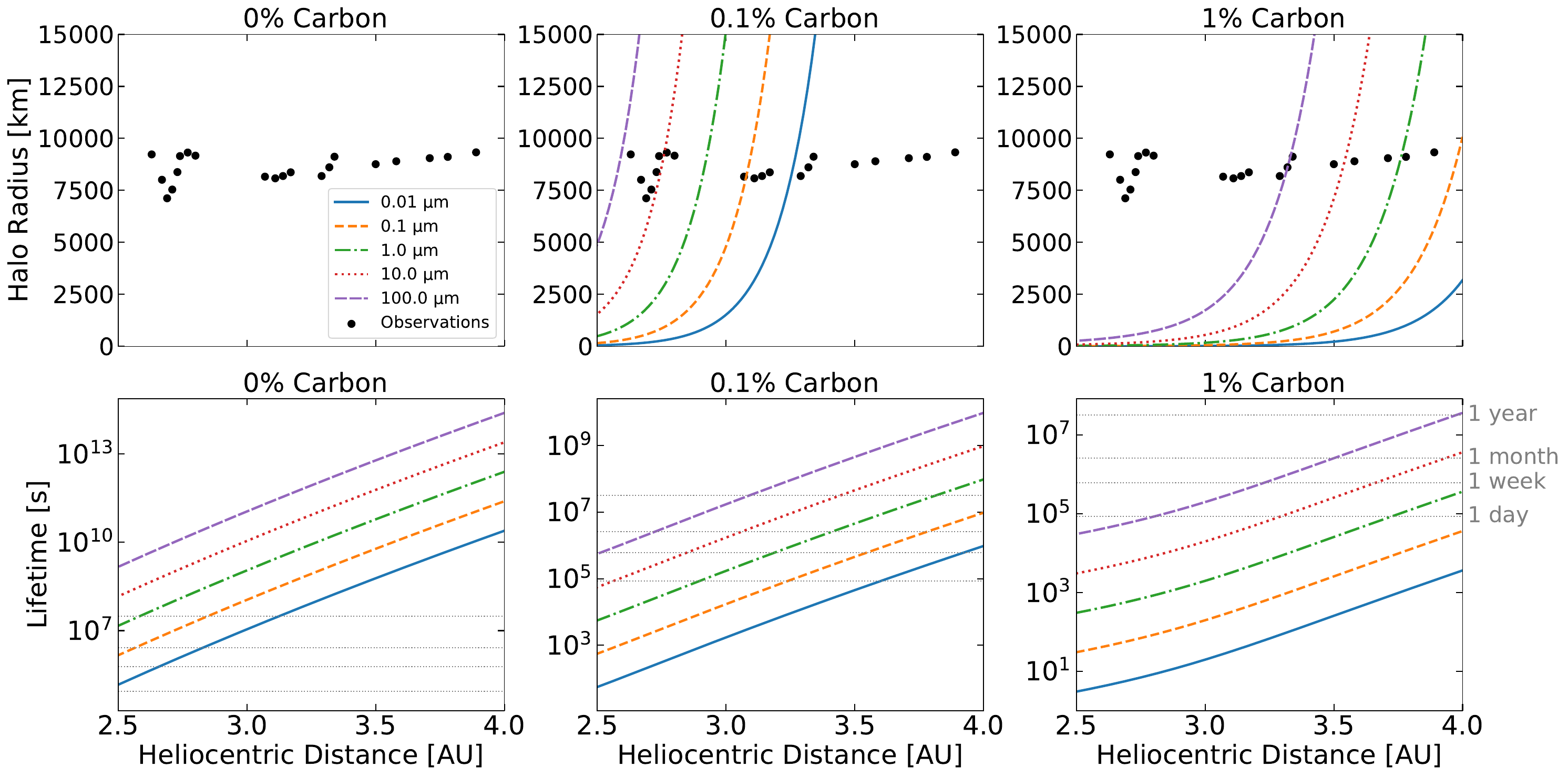}
	\caption{Model halo radius and grain lifetime of icy grains as functions of heliocentric distance for three assumed carbon fractions. The upper row shows the predicted halo radius for carbon fractions of 0\%, 0.1\%, and 1\%, while the lower row shows the corresponding grain lifetime on a logarithmic scale. In each panel, different line styles and colors denote grain radii of 0.01, 0.1, 1, 10, and \(100~\mu\mathrm{m}\), as indicated in the legend. In the upper panels, the black points show the observed halo radii. In the lower panels, the horizontal dotted lines mark reference timescales of 1 day, 1 week, 1 month, and 1 year.}
	\label{lifetime}
\end{figure*}

where \(r_{\rm h}\) is the heliocentric distance in au, \(S_\odot\) is the solar flux at 1 au and \(B_\lambda\) is the Planck function. The quantities \(L_v\) and \(Z\) are the latent heat of sublimation and the sublimation mass flux per unit grain surface area. Figure~\ref{temp} shows the resulting equilibrium temperatures at \(r_{\rm h}=3.0\) au for pure water-ice grains and for ice grains containing 1\% silicate or 1\% carbon by volume. The pure-ice and silicate-bearing grains have nearly identical temperatures over the full range of radii. The close overlap of these two curves indicates that a 1\% abundance of weakly absorbing silicate has little effect on the thermal balance. In contrast, the grains containing 1\% carbon are much warmer, with equilibrium temperatures of about 155 to 170 K and only a weak dependence on radius. This difference reflects the larger optical absorption of carbonaceous material at solar wavelengths. Even a small carbonaceous component therefore strongly increases the grain temperature. Because the vapor pressure of water ice is exponentially sensitive to temperature, this temperature increase leads to a much larger sublimation-driven radius-loss rate.

For each adopted \(a\), \(r_{\rm h}\), and composition, the energy balance equation was solved numerically for \(T\). The radius-loss rate was then computed from \(T\) following \citet{moreno2018icy}. We write the result as a positive loss rate,
\begin{equation}
    \left|\frac{da}{dt}\right|
    =
    2.28\times10^{4}
    T^{7/2}
    \exp\left[
        -\frac{
        5505+165(T/140)^{2.4}
        }{T}
    \right],
    \label{eq:adot_loss}
\end{equation}
where $|\frac{da}{dt}|$ is expressed in $\mu{\rm m~s^{-1}}$. Figure~\ref{sizeloss} shows the resulting radius-loss rate for grains with \(a=1~\mu{\rm m}\) as a function of heliocentric distance. The loss rate decreases rapidly with increasing \(r_{\rm h}\), reflecting the exponential dependence of the water-ice vapor pressure on temperature. The pure-ice and 1\% silicate curves are nearly indistinguishable over the plotted range, indicating that this amount of weakly absorbing silicate has little effect on the thermal balance. In contrast, the addition of 1\% carbon produces much larger loss rates at all heliocentric distances. This behavior follows from the higher optical absorption of carbonaceous material, which raises the equilibrium grain temperature and greatly enhances sublimation.

\subsection{Sublimation lifetime}
Following \citet{stevenson2012near}, we approximate the observed halo radius as the characteristic distance over which an icy grain survives in flight before complete sublimation. In this approximation, the halo radius is written as
\begin{equation}
L_{\rm halo}(a,r_{\rm h},f_{\rm C}) = v(a)\,\tau(a,r_{\rm h},f_{\rm C}),
\end{equation}
where \(v(a)\) is the grain velocity, \(\tau\) is the grain lifetime, and \(f_{\rm C}\) denotes the carbon fraction. As a simplifying assumption, and following \citet{liu2025great}, we neglected the heliocentric-distance dependence of the grain velocity and adopted a size-dependent velocity of $v(a)=v_0\left({a}/{1~\mu{\rm m}}\right)^{-1/2}$, with $v_0=88~{\rm m\,s^{-1}}$. The grain lifetime was defined as
\begin{equation}
\tau=\frac{a}{|da/dt|},
\end{equation}
where \(a\) is the grain radius and \(|da/dt|\) is the sublimation-driven radius-loss rate given by Eq.~\ref{eq:adot_loss}. Using this scheme, we calculated model halo radii for grain sizes of 0.01, 0.1, 1, 10, and \(100~\mu{\rm m}\) and for carbon fractions of 0, 0.1\%, and 1\% over the heliocentric distance range \(r_{\rm h}=2.4\)--4.0 au, which spans the observed interval. The resulting model halo radii and grain lifetimes are shown in Fig.~\ref{lifetime}.

Over the explored parameter space, Fig.~\ref{lifetime} shows that the model halo radius increases monotonically with heliocentric distance. This behavior is found for all grain sizes and carbon fractions and reflects the corresponding increase in grain lifetime at larger \(r_{\rm h}\), where weaker solar heating reduces the sublimation rate. As shown in the lower panels, the lifetime depends strongly on both grain size and carbon content. Larger grains and grains with lower carbon fractions survive longer, whereas smaller or more carbon-rich grains sublimate more rapidly. Consistent with previous studies showing that even minor refractory contamination can reduce the survival time of icy grains, our calculations indicate that a small carbon admixture can shorten the grain lifetime. For example, at \(r_{\rm h}=3\) au, the model lifetime of a \(10~\mu{\rm m}\) pure-ice grain is \(\sim 1.1\times10^{10}\) s, whereas it decreases to \(1.7\times10^{6}\) s and \(2.0\times10^{5}\) s for grains containing 0.1\% and 1\% carbon, respectively.

Compared with the observations, the pure-ice models predict halo radii larger than observed because pure-ice grains survive too long under the adopted sublimation rates. Adding a small carbon fraction shortens the grain lifetime and reduces the model halo radius for a given grain size, bringing the model scale closer to the observations. However, the observed halo radius does not vary monotonically with heliocentric distance, but instead shows changes associated with the activity enhancements discussed in Sect.~\ref{sec:halo}. Therefore, no single fixed combination of grain size and carbon fraction reproduces the full set of measurements. Within the fixed-composition parameter space explored, icy grains with radii of $10~\mu{\rm m}$ and a carbon fraction of 0.1\% yield model halo radii comparable to the observed halo scale during the enhanced-activity interval from 21 to 25 February 2025. For this parameter combination, the model grain lifetime varies from $2.8\times10^5$ to $3.2\times10^5$ s. Under the adopted velocity scaling, the corresponding propagation distance is about $7800$--$8900$ km, comparable to the measured halo radius during this interval. This result suggests that slightly carbon-bearing icy grains of this size can account for the observed halo scale during the late-February activity enhancement.

\begin{figure}
	\centering
	\includegraphics[width=\columnwidth]{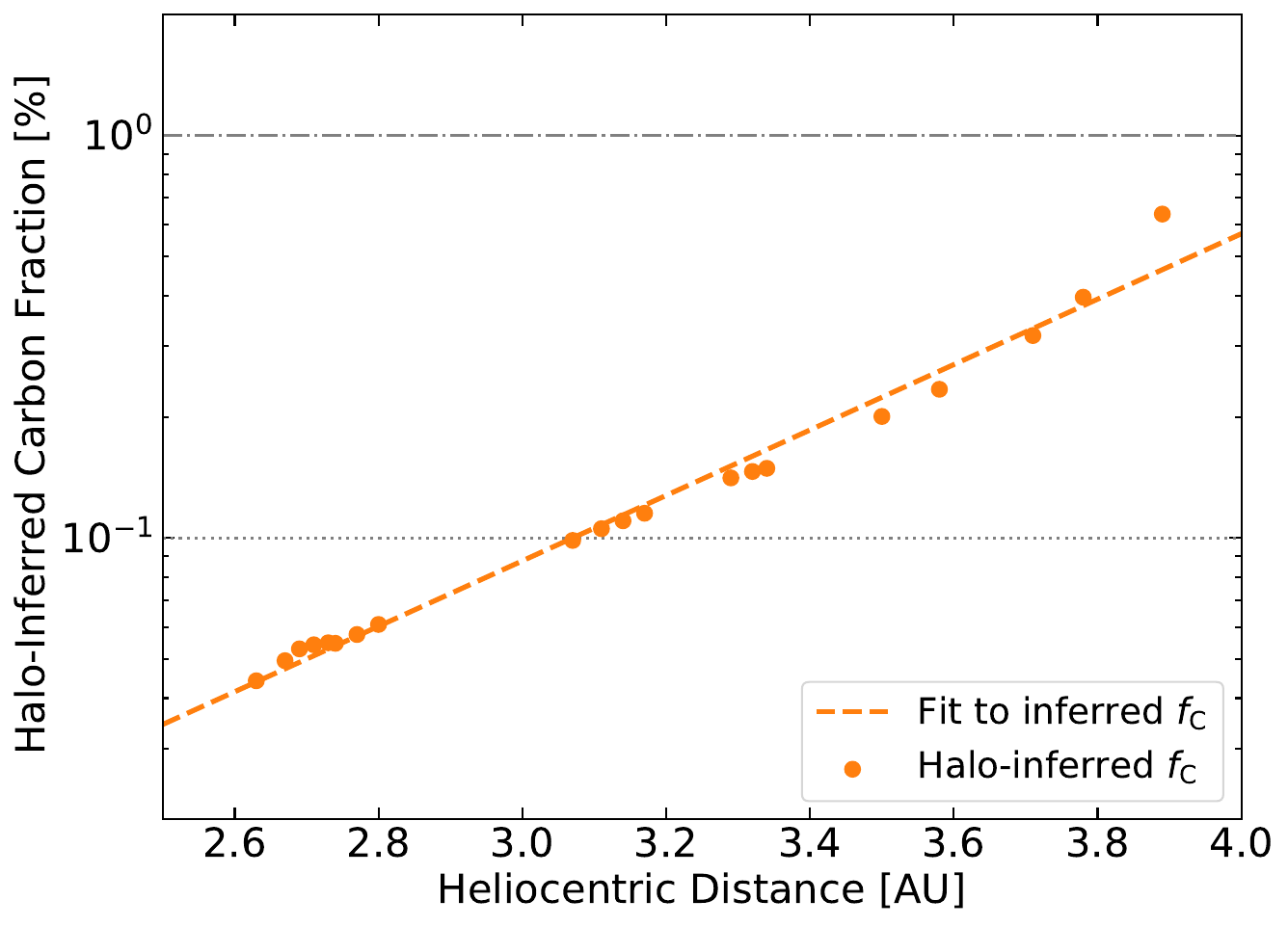}
	\caption{Halo-inferred carbon fraction as a function of heliocentric distance for a fixed grain radius of $a = 0.1~\mu\mathrm{m}$. The dashed line shows a log-linear fit to the inferred fractions. The horizontal dotted and dash-dotted lines indicate carbon fractions of $0.1\%$ and $1\%$, respectively.}
	\label{fig:variable_carbon}
\end{figure}

Because the fixed-composition models provide only local rather than global agreement with the observations, we next examined whether composition changes alone could reproduce the broader halo evolution. As an illustrative case, we selected a fixed grain radius of $a = 0.1~\mu\mathrm{m}$. The carbon fraction was then allowed to vary between observing epochs, and at each epoch we inverted the thermal model to determine the value required for the modeled propagation distance to match the measured halo radius. The resulting halo-inferred carbon fractions increase approximately log-linearly with heliocentric distance, from about $0.05\%$ at $r_{\mathrm{h}} \sim 2.6~\mathrm{au}$ to about $0.6\%$ near $r_{\mathrm{h}} \sim 3.9~\mathrm{au}$ (Fig.~\ref{fig:variable_carbon}). This calculation shows that, at least mathematically, the broad halo-radius evolution can be reproduced by allowing the composition to vary at fixed grain size.

A possible physical interpretation of the inferred composition trend is that
	grains ejected at different epochs represent distinct populations released from
	an evolving near-surface layer. Progressive sublimation and depletion of water
	ice during the post-perihelion evolution may leave the active near-surface
	material increasingly enriched in refractory components. Consequently, grains
	released at later epochs may have had a higher population-averaged carbon
	fraction than those released earlier, providing a plausible explanation for the
	broad halo-radius evolution. The short-term, non-monotonic variations observed around the photometric
	enhancements may reflect episodic releases of fresh icy grains superimposed on
	this longer-term compositional trend. This interpretation, however, is conditional on the assumption of a fixed
	grain radius. Temporal changes in the characteristic grain size, the active source regions, or the
	relative contributions of different grain populations could produce a similar
	halo-radius evolution. In the absence of independent measurements, the variable-composition scenario should therefore be regarded as a plausible, but not unique, physical interpretation.

\section{Conclusions}
\label{CONCLUSIONS}

Our analysis of archival ZTF images of comet C/2023~A3 (Tsuchinshan--ATLAS) during its post-perihelion phase leads to the following conclusions.

\begin{enumerate}
\item The comet underwent a secular fading over the observing interval. Within a fixed projected aperture radius of 40,000 km, the apparent $r$-band magnitude changed from $m_r = 12.86$ on 15 February 2025 to $m_r = 14.77$ on 26 May 2025, as the heliocentric distance increased from 2.64 to 3.90 au. Two short-lived brightenings were superimposed on this overall decline, occurring in late February and early April.
      
\item The effective dust scattering cross-section decreased by a factor of about 2.4 and followed $C_{\rm e} \propto r_{\rm h}^{-3.1}$, while the dust mass production rate scaled as $\dot{M}_{\rm d} \propto r_{\rm h}^{-3.41}$. These trends indicate a gradual weakening of the post-perihelion dust activity and are compatible with a contribution from water-ice-driven sublimation. Extrapolating the fitted post-perihelion dust-production relation back to perihelion gives a model-dependent cumulative dust mass loss of $3.0 \times 10^{11}$ kg, corresponding to a globally averaged erosion depth of $\sim 1.4$ m for an assumed nucleus radius of 5.9 km and bulk density of 500 kg m$^{-3}$.
      
\item The azimuthally averaged surface-brightness profiles are consistently shallower than the canonical $1/\rho$ profile expected for a steady-state coma in free expansion. This behavior indicates radial evolution in the dust distribution or scattering properties. Delayed fragmentation of dust aggregates provides a plausible explanation.
      
\item A persistent central excess is present in the radial surface-brightness profiles. Forward modeling shows that a smooth coma plus an unresolved central source cannot fully reproduce this excess. Together with the less-red inner-coma color, this excess points to a compact near-nucleus component associated with fresh or ice-rich grains. The halo radius varied non-monotonically and showed local increases near the short-lived activity enhancements, consistent with episodic replenishment of near-nucleus icy grains.
      
\item Among the icy-grain models explored, grains with radii of $\sim 10\,\mu{\rm m}$ and a carbon fraction of $\sim 0.1\%$ yield model halo radii comparable to the observed halo scale during the enhanced-activity interval from 21 to 25 February 2025, with model lifetimes of $(2.8-3.2)\times10^5$ s. However, no single combination of fixed grain size and composition reproduces the full halo evolution. A complementary fixed-size test shows that an increase in the carbon fraction can reproduce the broad halo-radius trend, possibly reflecting progressive refractory enrichment of the near-surface material. The short-term, non-monotonic variations around the photometric enhancements remain consistent with episodic releases of fresh icy grains. 
 
\end{enumerate}

\begin{acknowledgements}
The work is supported by the Science and Technology Development Fund (FDCT) of Macau (grant Nos. 0008/2024/AKP, 0034/2024/AMJ, 0008/2024/ITP1, 002/2024/SKL, 0002/2025/AKP) and the National Key Research and Development Program of China (grant No. 2024YFE0201000) to X. Zhang, and the National Natural Science Foundation of China (No.~12472048) to X. Liu. This work is based on observations obtained with the Samuel Oschin 48-inch Telescope at Palomar Observatory as part of the Zwicky Transient Facility (ZTF). ZTF is supported by the National Science Foundation under grants Nos. AST-1440341 and AST-2034437. Operations are conducted by COO, IPAC, and UW.
\end{acknowledgements}
%%%%%%%%%%%%%%%%%%%%%%%%%%%%%%%%%%%%%%%%%%%%%%%%%%.

\bibliography{aanda}{}
\bibliographystyle{aa}

\clearpage
\onecolumn

\begin{appendix} 
	
\section{\mbox{Photometric and dust properties of C/2023 A3}} 
\label{appdA}
\begin{table*}[!htbp]
	\centering
	\renewcommand{\arraystretch}{1.4}
	\caption{Photometric and dust properties of comet A3 derived from the ZTF observations.}
	\label{tab:photometry_dust}
	\setlength{\tabcolsep}{18pt}
	\small
	
	\begin{tabular}{lccccc}
		\toprule
		Date & $m_r^a$ & $H_r^b$ & $C_{\rm e}^c$ & $M_{\rm d}^d$ & $\dot{M}_{\rm d}^e$ \\
		(UT) & (mag) & (mag) & ($10^3$ km$^2$) & ($10^9$ kg) & (kg s$^{-1}$) \\
		\midrule
		15 February 2025 & $12.862 \pm 0.009$ & $7.75 \pm 0.05$ & $9.68 \pm 0.78$ & $3.87 \pm 0.32$ & $326 \pm 27$ \\
		18 February 2025 & $12.926 \pm 0.012$ & $7.76 \pm 0.05$ & $9.61 \pm 0.46$ & $3.84 \pm 0.18$ & $323 \pm 16$ \\
		19 February 2025 & $13.098 \pm 0.008$ & $7.92 \pm 0.05$ & $8.34 \pm 0.39$ & $3.33 \pm 0.17$ & $280 \pm 13$ \\
		21 February 2025 & $13.085 \pm 0.007$ & $7.87 \pm 0.05$ & $8.72 \pm 0.42$ & $3.50 \pm 0.17$ & $291 \pm 13$ \\
		22 February 2025 & $12.931 \pm 0.010$ & $7.69 \pm 0.05$ & $10.22 \pm 0.50$ & $4.09 \pm 0.20$ & $340 \pm 17$ \\
		23 February 2025 & $13.064 \pm 0.010$ & $7.81 \pm 0.05$ & $9.19 \pm 0.45$ & $3.68 \pm 0.18$ & $306 \pm 15$ \\
		25 February 2025 & $13.006 \pm 0.005$ & $7.72 \pm 0.05$ & $10.01 \pm 0.49$ & $4.00 \pm 0.20$ & $332 \pm 16$ \\
		27 February 2025 & $13.247 \pm 0.007$ & $7.93 \pm 0.05$ & $8.26 \pm 0.41$ & $3.30 \pm 0.17$ & $272 \pm 13$ \\
		20 March 2025 & $13.740 \pm 0.013$ & $8.12 \pm 0.06$ & $6.92 \pm 0.40$ & $2.76 \pm 0.17$ & $218 \pm 12$ \\
		23 March 2025 & $13.921 \pm 0.010$ & $8.27 \pm 0.06$ & $6.05 \pm 0.35$ & $2.42 \pm 0.14$ & $189 \pm 11$ \\
		25 March 2025 & $13.855 \pm 0.010$ & $8.18 \pm 0.06$ & $6.56 \pm 0.39$ & $2.62 \pm 0.15$ & $205 \pm 12$ \\
		27 March 2025 & $14.045 \pm 0.030$ & $8.35 \pm 0.07$ & $5.62 \pm 0.55$ & $2.25 \pm 0.22$ & $174 \pm 11$ \\
		6 April 2025  & $14.276 \pm 0.012$ & $8.48 \pm 0.07$ & $4.95 \pm 0.30$ & $1.98 \pm 0.12$ & $151 \pm 10$ \\
		8 April 2025  & $14.182 \pm 0.020$ & $8.37 \pm 0.07$ & $5.48 \pm 0.35$ & $2.19 \pm 0.14$ & $167 \pm 11$ \\
		10 April 2025 & $14.152 \pm 0.019$ & $8.33 \pm 0.07$ & $5.72 \pm 0.36$ & $2.29 \pm 0.15$ & $173 \pm 11$ \\
		23 April 2025 & $14.593 \pm 0.013$ & $8.68 \pm 0.07$ & $4.13 \pm 0.26$ & $1.65 \pm 0.11$ & $121 \pm 7$ \\
		29 April 2025 & $14.738 \pm 0.034$ & $8.79 \pm 0.07$ & $3.72 \pm 0.25$ & $1.48 \pm 0.11$ & $109 \pm 7$ \\
		10 May 2025 & $14.800 \pm 0.026$ & $8.81 \pm 0.07$ & $3.68 \pm 0.23$ & $1.47 \pm 0.09$ & $105 \pm 6$ \\
		16 May 2025 & $14.805 \pm 0.024$ & $8.79 \pm 0.07$ & $3.73 \pm 0.23$ & $1.50 \pm 0.09$ & $107 \pm 6$ \\
		26 May 2025 & $14.769 \pm 0.012$ & $8.72 \pm 0.06$ & $3.99 \pm 0.21$ & $1.60 \pm 0.09$ & $111 \pm 6$ \\
		\bottomrule
		\multicolumn{6}{l}{$^a$ Apparent magnitude measured in the ZTF $r$-band.} \\
		\multicolumn{6}{l}{$^b$ Absolute magnitude.} \\
		\multicolumn{6}{l}{$^c$ Effective scattering cross-section area.} \\
		\multicolumn{6}{l}{$^d$ Dust mass within the aperture.} \\
		\multicolumn{6}{l}{$^e$ Dust mass production rate.} \\
	\end{tabular}
\end{table*}

\end{appendix}

\end{document}